\documentclass[12pt]{article}
\usepackage{epsfig}
\usepackage[utf8]{inputenc}
\usepackage{color}

\definecolor{grey}{rgb}{0.4,0.4,0.4}

\def\be{\begin{equation}}
\def\ee{\end{equation}}
\def\bea{\begin{eqnarray}}
\def\eea{\end{eqnarray}}

\begin{document}

\title{The $W_LW_L$ scattering at the LHC: improving the selection
  criteria}

\author{Krzysztof Doroba$^a$, Jan Kalinowski$^{a,b}$, Jakub
  Kuczmarski$^a$, \\ Stefan Pokorski$^a$, Janusz Rosiek$^a$, Micha{\l}
  Szleper$^c$, \\ S{\l}awomir Tkaczyk$^d$ \and
  \normalsize \sl $^a$Physics Department, University of Warsaw,
  Ho{\.z}a 69, 00-681 Warsaw, Polandy\\[1pt] \and
  \normalsize \sl $^b$University of Hamburg, Luruper Chaussee 149,
  D-22761  Hamburg, Germany,\\[-4pt]
  \normalsize \sl DESY, Notkestrasse 85, D-22607 Hamburg,
  Germany\\[1pt] \and
  \normalsize \sl $^c$National Center for Nuclear Research, High
  Energy Physics Department, \\[-4pt] \normalsize \sl Ho{\.z}a 69,
  00-681, Warszawa, Poland \and
  \normalsize \sl $^d$Fermi National Accelarator Laboratory, Batavia,
  IL 60510, USA}

\date{\today}

\maketitle

\begin{abstract}
  We present a systematic study of the different mechanisms leading to
  $WW$ pair production at the LHC, both in the same-sign and
  opposite-sign channels, and we emphasize that the former offers much
  better potential for investigating non-resonant $W_LW_L$ scattering.
  We propose a new kinematic variable to isolate the $W_LW_L$
  scattering component in same-sign $WW$ production at the LHC.
  Focusing on purely leptonic $W$ decay channels, we show that it
  considerably improves the LHC capabilities to shed light on the
  electroweak symmetry breaking mechanism after collecting $100
  fb^{-1}$ of data at $\sqrt{s} = 14$ TeV.  The new variable is less
  effective in the opposite-sign $WW$ channel due to different
  background composition.
\end{abstract}

\section{Introduction}
\label{sec:intro}

The longitudinal $WW$ scattering carries the most direct information
about the mechanism of electroweak symmetry breaking, no matter
whether a physical elementary Higgs particle exists or some kind of
strongly interacting physics is responsible for this breaking.  In
fact, even if a light Higgs boson is discovered, the energy dependence
of the longitudinal $WW$ scattering above the Higgs mass scale will
tell us if the Higgs boson unitarizes the $WW$ scattering fully or
only partially, as in some theoretical models with composite
Higgs~\cite{partial}.  Experimental investigation of the $W_LW_L$
scattering as a function of its center-of-mass energy ${M_{WW}}$
becomes feasible at the LHC.  The techniques for observing the
$W_LW_L$ scattering signal in $pp$ collisions have been extensively
investigated and reported in many papers
(\cite{bagger}--\cite{chiara}).  They are based on the differences in
the emission process of the transverse and longitudinal gauge bosons
from the colliding quarks and in the behavior of the $WW$ scattering
amplitudes as a function of their center-of-mass energy and the
scattering angle.  Those effects are, however, strongly masked by the
quark distribution functions inside the proton and by various sources
of large background.  It has been found that such techniques as
forward jet tagging, central jet vetoing, and cuts on the final lepton
transverse momenta are very promising in the isolation of the $W_LW_L$
scattering signal from the background.  At the same time, the above
studies clearly showed that such measurement would be experimentally
very challenging and typically require from more than a year to
several years of LHC running at full nominal parameters to obtain
observable effects.  In view of this, a deeper understanding of the
process on the theoretical side and possible new techniques allowing
for a better event selection and a better data analysis can play an
important role to make the study feasible in a relatively shorter
timescale.

Given the importance of the experimental access to the longitudinal
$WW$ scattering channel, we readdress this issue in the present paper.
We analyze the feasibility of observing the signal of an enhanced,
compared to the prediction of the SM with a light Higgs boson,
$W_LW_L$ scattering at the LHC running at an energy of 7, 8 and 14
TeV.  As our laboratory for the enhanced $W_LW_L$ scattering we use
the higgsless scenario.  If a Higgs boson does exist, but with
couplings modified compared to the SM so that it does not fully
unitarize the $WW$ scattering, the signal will be weaker and more
difficult to observe (see e.g.~\cite{grojean}).

In search for the most efficient event selection criteria, we first
present a systematic study of the different mechanisms leading to $WW$
pair production at the LHC, both in the same-sign and opposite-sign
channels.  We justify why and under which conditions the complicated
process of $WW$ production in $pp$ collisions can be reduced to
general considerations on $W_L$ and $W_T$ emission off a quark line.
We then examine the differences between emission and scattering
processes of the $W$ bosons of well defined, longitudinal or
transverse polarizations.  For quantitative analyses we apply exact
matrix element calculations involving the MADGRAPH
generator~\cite{madgraph}.  We also briefly show that some basic
features of $W$ emission can be qualitatively understood within the
framework of the Effective $W$ Approximation (EWA)~\cite{EWA}.

Based on our study, we propose a new kinematic variable for the
longitudinal $W^+W^+$ scattering signal, showing that it gives
significant improvement of the signal-to-background ($S/B$) figures
compared to all the previously discussed selection criteria.  We
furthermore emphasize that $WW$ scattering with same-sign $W$'s offers
the best physics potential for the LHC.

In this work, we choose to focus on the purely leptonic $W$ decays in
both the same-sign and opposite-sign $WW$ scattering processes.  These
channels are known as ``gold plated", in spite of their low
statistics, for their relatively low background contaminations and
experimental purity.

\section{Signal and background definitions}
\label{sec:sbdef}

Following the well established ``subtraction method"~\cite{bagger}, we
define the signal as the enhancement in the production of $WW$ pairs
in association with two parton jets in the higgsless scenario over the
prediction of the Standard Model with a 120 GeV Higgs:
\bea
\mathrm{Signal} = \sigma(pp \rightarrow jjWW)|_{\rm higgsless} -
\sigma(pp \rightarrow jjWW)|_{\rm M_h=120GeV}.
\label{eq:sigdef0}\nonumber\\
\eea
It is worth to note that even an observation of a light Higgs boson
does not necessarily exclude a non-vanishing signal, assuming that it
may have non-standard couplings to vector bosons.  The signal comes on
top of the irreducible background, which is:
\bea
\mathrm{Background} = \sigma(pp \rightarrow jjWW)|_{\rm M_h=120GeV}.
\eea
For the calculations in the higgsless scenario, we make no distinction
between removing all Feynman diagrams involving Higgs exchange or
setting an inaccessibly large Higgs mass ($10^{10}$ GeV) in the
calculation.  Furthermore, instead of using any particular
unitarization model, we will assume the $W_LW_L$ scattering cross
section saturates just before its partial wave amplitudes hit the
perturbative unitarity limit at a $WW$ center of mass energy around
1.2 TeV and stays constant above this value.  Such assumption defines
the practical limiting case for a signal size which is still
consistent with all physics principles.  Once all the appropriate
selection criteria are applied, we find this prescription reduce the
signal size by 20-23\%, depending on the selection details, compared
to the purely mathematical higgsless case with no unitarization at all
(of course, mostly affected is the most interesting region of large
invariant $WW$ masses).

To a good approximation, the signal comes from the scattering of
longitudinally polarized $W$ bosons, by far the most sensitive to the
electroweak symmetry breaking dynamics.  Irreducible background is
dominated by the transverse $W$ production.

The cross sections appearing in the definition of the signal and the
background must be calculated in a gauge invariant way, including also
$O(\alpha^2)$ (pure EW) and $O(\alpha\alpha_S)$ (QCD) diagrams in
which two $W$'s are produced but do not interact.  As we shall discuss
below, QCD effects in particular play an important role in the correct
calculation of the background.

\begin{figure}[htbp]
\begin{tabular}{ll}
  \epsfig{file=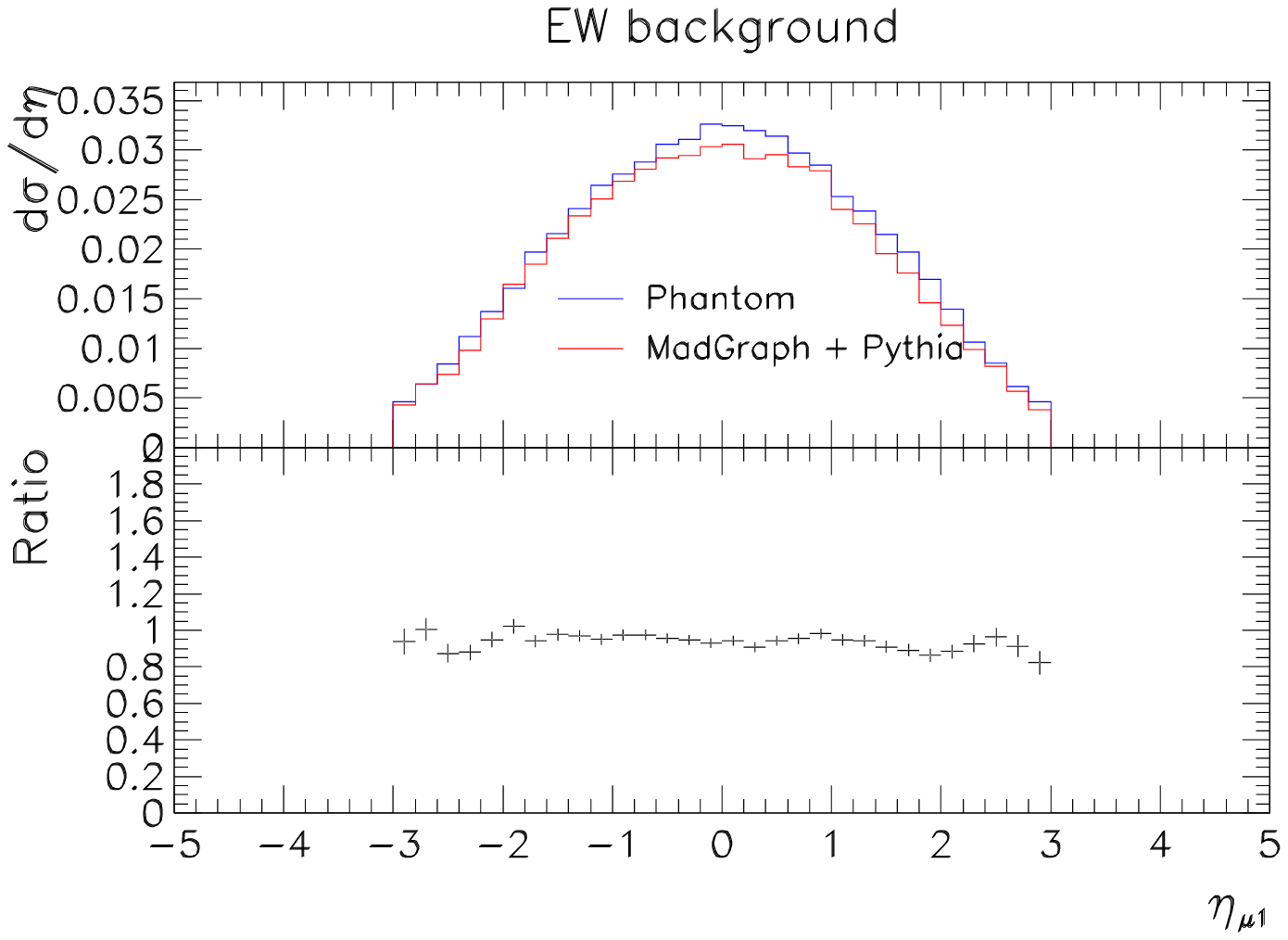,width=0.49\linewidth} &
  \epsfig{file=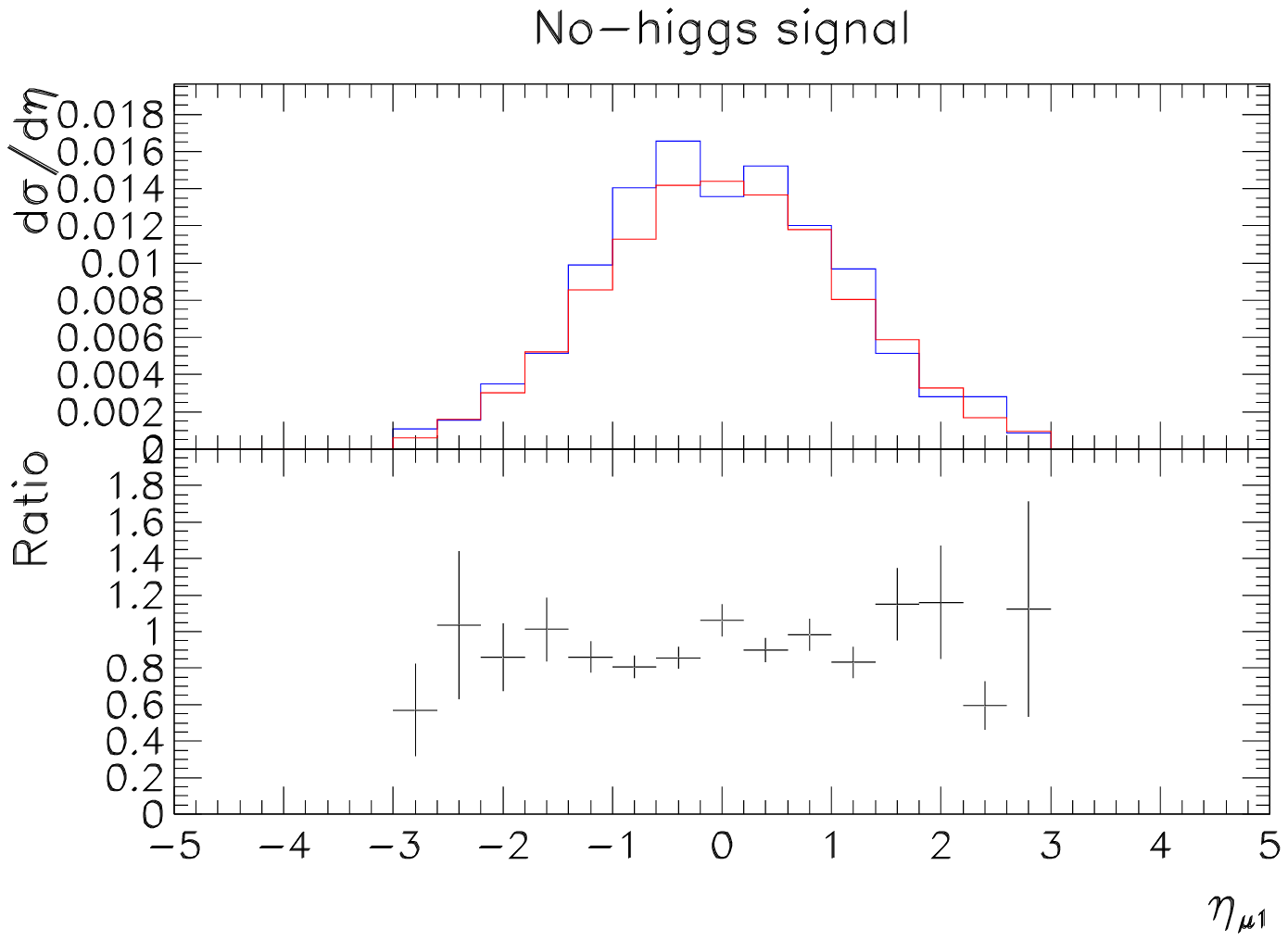,width=0.49\linewidth}
  \\
  a) & b) \\
  \epsfig{file=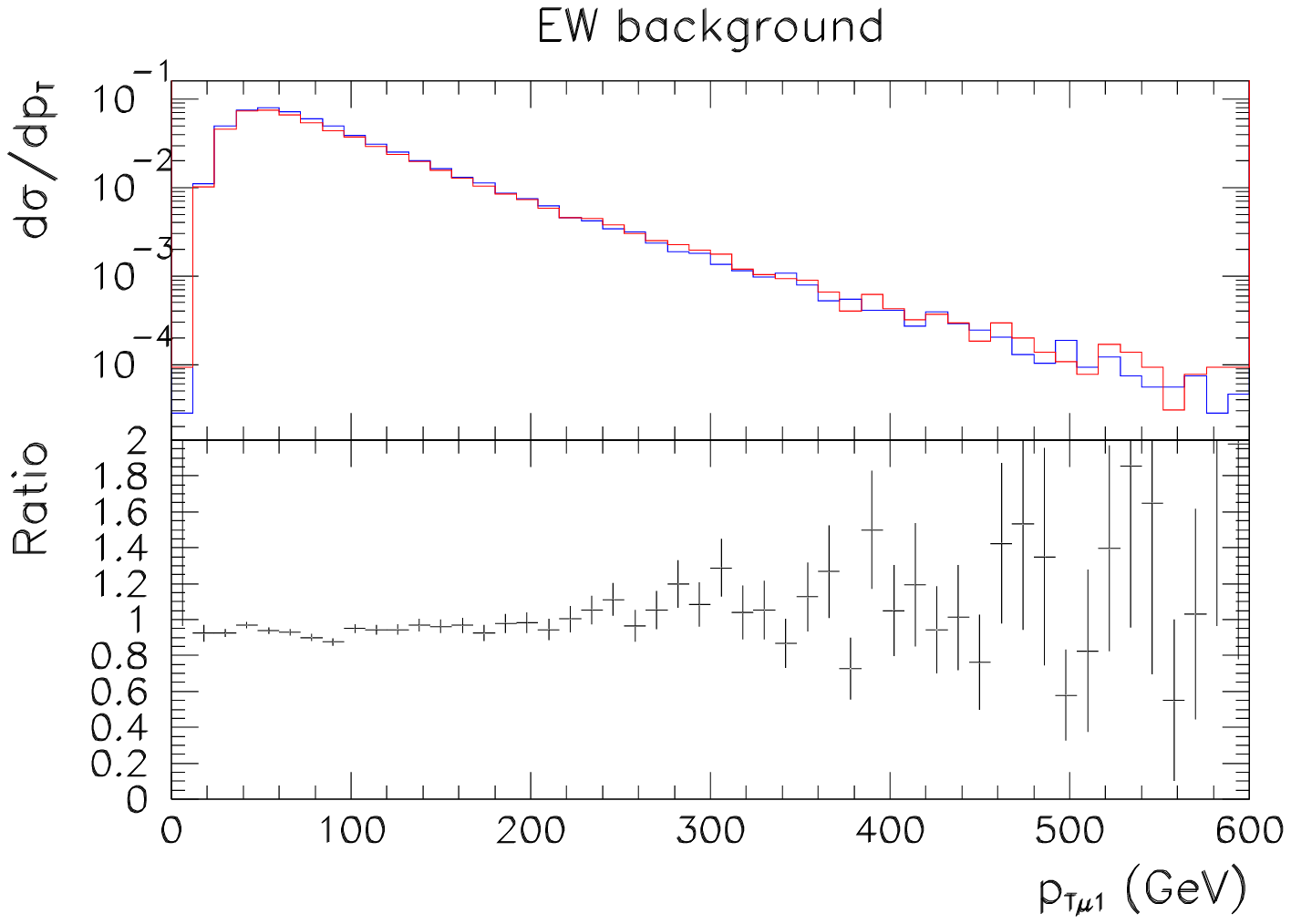,width=0.49\linewidth} &
  \epsfig{file=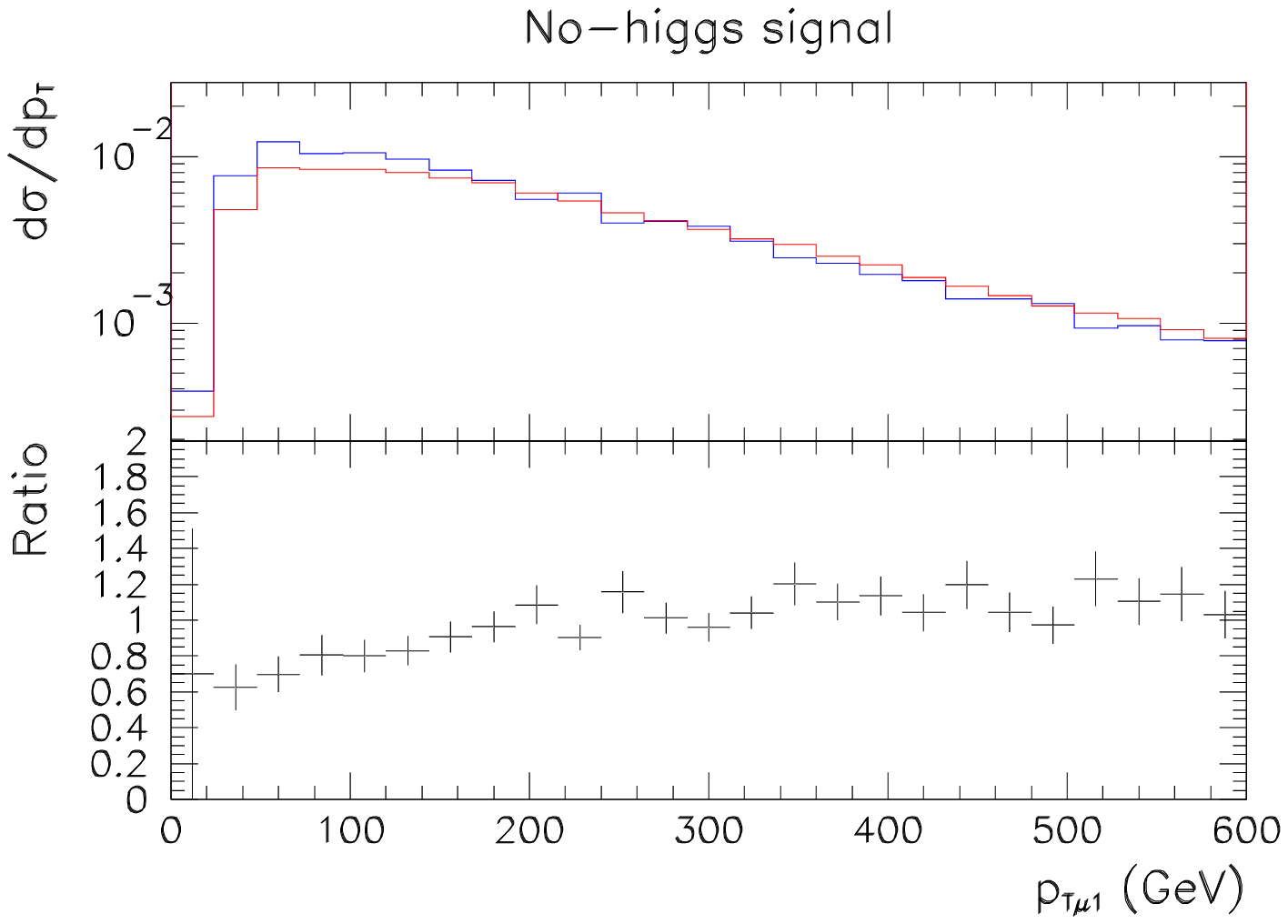,width=0.49\linewidth}
  \\ 
  c) & d) \\
  \epsfig{file=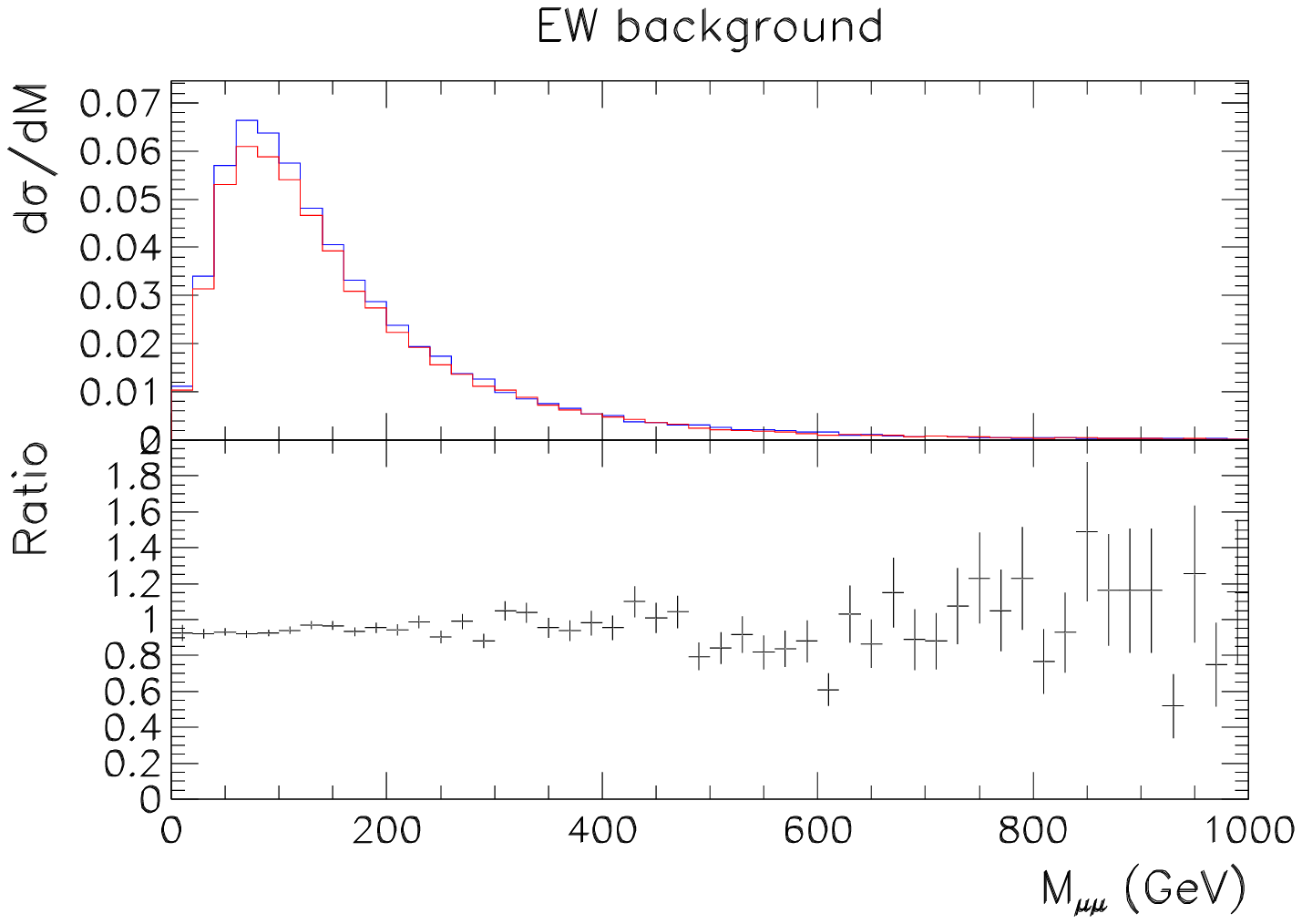,width=0.49\linewidth} &
  \epsfig{file=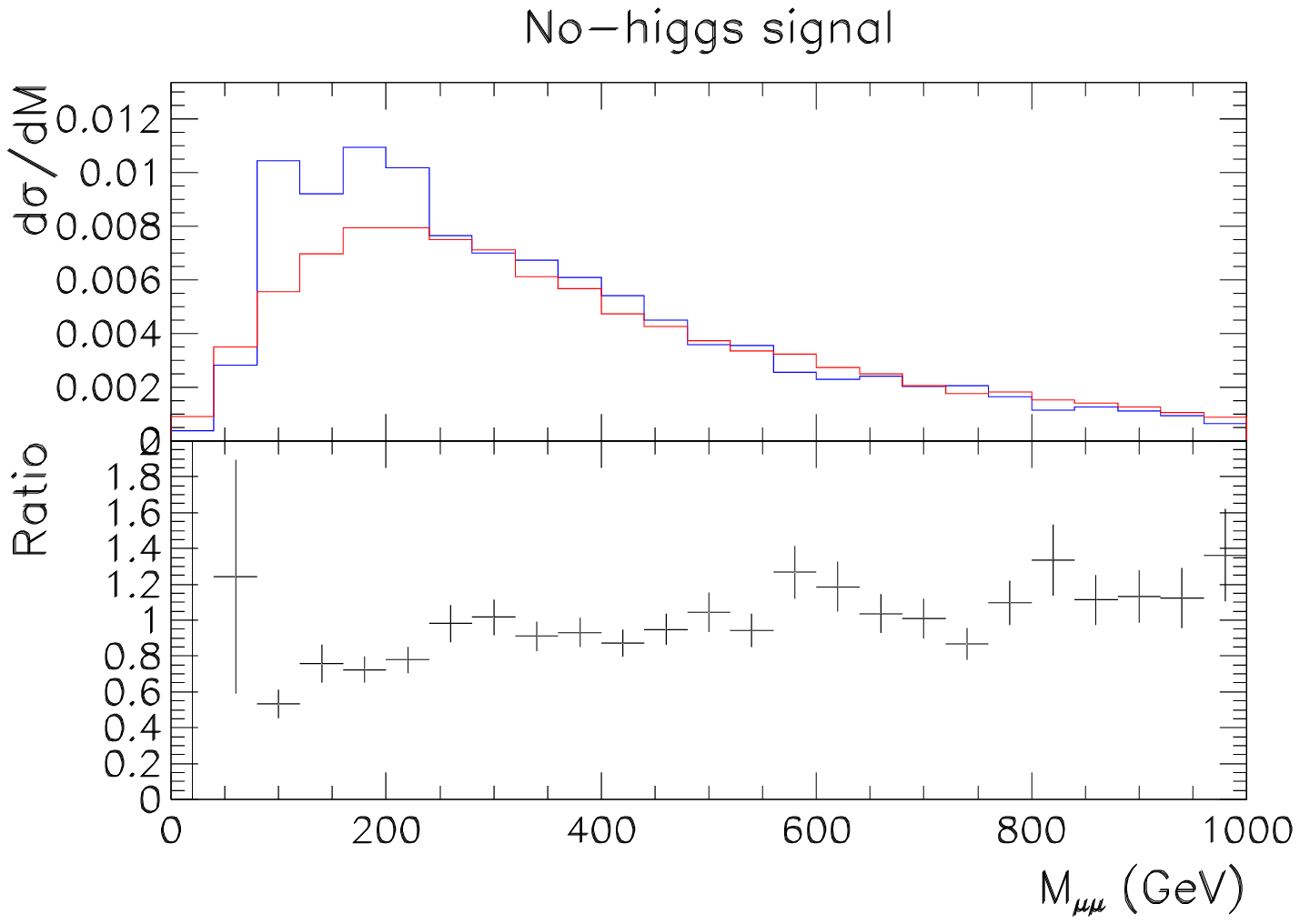,width=0.49\linewidth} 
  \\
  e) & f) 
\end{tabular}
\caption{Kinematic distributions of final state muons from the $pp
  \rightarrow jj\mu^+\mu^+\nu\nu$ process at 14 TeV, obtained using
  the $W$ on-shell approximation (labeled MadGraph+Pythia) and exact
  matrix element calculations (labeled Phantom).  Pseudorapidities
  (a,b), transverse momenta (c,d) and invariant masses (e,f) are
  shown.  For the sake of comparison, background samples contain pure
  EW processes only and a 200 GeV Higgs is assumed.  }
\label{fig:prodxdecay}
\end{figure}

Both for the signal and for the background we perform a full matrix
element calculation of the processes $pp \rightarrow jjW^+W^+$ and $pp
\rightarrow jjW^+W^-$ using the MADGRAPH generator, followed by
leptonic $W$ decay and quark hadronization.  The procedure implies
that the intermediate state $W$'s are on-shell, which in literature is
sometimes referred to as the ``production $\times$ decay"
approximation.  In particular, diagrams which contribute to the same
final state but do not proceed via an intermediate state consisting of
a $WW$ pair are here neglected.  The $W$ decay is simulated using a
private version of PYTHIA~6 which has the appropriate angular
distributions for the decays of $W_L$ and $W_T$ implemented.  Such
modification of the code (in the standard distribution of
PYTHIA~6~\cite{pythia} the $W$'s are decayed isotropically) is vital
in order to obtain the correct kinematic distributions of the final
state leptons.  To check the validity of the $W$ on-shell
approximation, a dedicated study was carried out, in which special
samples of the $pp \rightarrow jjW^+W^+ \rightarrow jjl^+l^+\nu\nu$
process generated using this method were compared in detail with $pp
\rightarrow jjl^+l^+\nu\nu$ samples generated with the PHANTOM
program~\cite{phantom}, without resorting to any ``production $\times$
decay" approximation\footnote{MS is grateful to the authors of
  Ref.~\cite{chiara} for granting him access to the PHANTOM data
  samples.}.  After imposing exactly the same generation cuts on both
samples, comparisons of the respective kinematic distributions of the
final state leptons do not reveal large differences, as shown in
Fig.\,\ref{fig:prodxdecay}.  An overall uncertainty in the
normalization amounting to a few per cent is apparent for the EW
background, but all lepton kinematics is adequately described.  The
signal calculated in the $W$ on-shell approximation reveals a deficit
of events at low invariant masses of the $l^+l^+$ pair ($M_{ll} < 200$
GeV).  However, this kinematic region is of little relevance for the
present analysis and, as we will see later, will be almost completely
eliminated by the selection criteria.  We conclude that the
``production $\times$ decay" approximation is unlikely to be a source
of any significant bias in the results.

We have also cross checked our calculations for the pure EW $jjW^+W^-$
production against the numbers from Table 7 of Ref.~\cite{zeppenfeld}
that were calculated using VBFNLO and including NLO QCD corrections to
the purely electroweak tree-level process.  At the level of their
``inclusive" cuts, for both Higgs boson mass hypotheses of 100 GeV and
1 TeV, we find satisfactory agreement, although the agreement in the
total cross section expectedly breaks down in the Higgs resonance
region.

In the ``production $\times$ decay" approximation, the signal and the
background can be calculated as respective sums over different
polarizations of the final $WW$ pairs.  The dependence of the
transverse and mixed transverse-longitudinal $WW$ pair production on
the Higgs boson mass, up to the higgsless limit, is weaker than 3\%
overall.  Although it is a part of the signal in the sense of
Eq.~(\ref{eq:sigdef0}) (and not necessarily negligible given $W_LW_L$
overall smallness), this contribution has no specific kinematic
signature that would allow to isolate it from the much larger
background.  In the kinematic range where $WW$ scattering can be
observed and background sufficiently suppressed it becomes completely
negligible and will be ignored further on.

Thus, our effective technical definitions of signal and background can
be rewritten as:
\bea
\mathrm{Signal} &=& \sigma(pp \rightarrow jjW_LW_L)|_{\rm higgsless} -
\sigma(pp \rightarrow jjW_LW_L)|_{\rm M_h=120GeV},
\label{eq:sigdef}\\
\mathrm{Background} &=& \sigma(pp \rightarrow jjW_LW_L)|_{\rm
  M_h=120GeV} + \sigma(pp \rightarrow jjW_TW_T) + \sigma(pp
\rightarrow jjW_TW_L),
\label{eq:bckgdef}\nonumber\\
\eea
where each polarization cross section is given in the most general
case by a sum of interfering amplitudes: pure electroweak diagrams
with interacting and non-interacting $WW$ pairs and QCD-electroweak
diagrams with non-interacting $WW$ pairs.

\section{$W$ pair production at the LHC.  Emission and scattering of
  longitudinal and transverse $W$'s}
\label{sec:prod+em}

Production of $WW$ pairs with two associated parton-level jets at the
LHC involves hundreds of tree-level diagrams and in general is
dominated by events with no direct relevance to the mechanism of
electroweak symmetry breaking.  Signal selection can be crudely
divided into two steps: ``basic'' cuts suppressing the soft
parton-parton collisions (both QCD and EW related) and leaving mostly
the events with hard $W$ interactions, then ``final'' cuts
distinguishing the $W_LW_L$ scattering from the irreducible background
mainly from $W_TW_T$ and $W_TW_L$ pairs.

The $WW$ production cross section with polarized $W$'s in the final
state can schematically be written as follows:
\bea
d\sigma_{ij}(M,m_h)=|a(M, m_h)_{ij}\alpha^2 +
b(M)_{ij}\alpha\alpha_s|^2
\label{eq:sigapp}
\eea
where $M$ is the invariant mass of the $WW$ pair, $m_h$ is either 120
GeV or denotes the higgsless case, $\alpha$ ($\alpha_s$) is the
electroweak (strong) coupling squared and subscripts $i$ and $j$
denote the final $W$ pair polarizations.  At the production amplitude
level the $a$ term sums all the pure EW contributions and the $b$ term
comes from a mixed EW/QCD process which is $m_h$ independent.  From
the above schematic formula the signal defined in
Eq.~(\ref{eq:sigdef}) can be written as
\bea
\mathrm{Signal} &\propto& \alpha^4\left[|a(M,
  \mathrm{higgsless})_{LL}|^2 -
  |a(M, m_h=120)_{LL}|^2\right] \nonumber\\
&+& 2 \alpha_s^2 \alpha^2 \mathrm{Re} \left[ \left(
    a(M,\mathrm{higgsless})_{LL} - a(M,m_h=120)_{LL} \right)
  b^{\star}(M)_{LL} \right]
\eea
From the above one sees that the signal is driven by pure EW
processes.  The large term $|b_{LL}|^2$ cancels out in the difference
through which we define the signal.
\begin{figure}[htb]
\begin{center}
  \epsfig{file=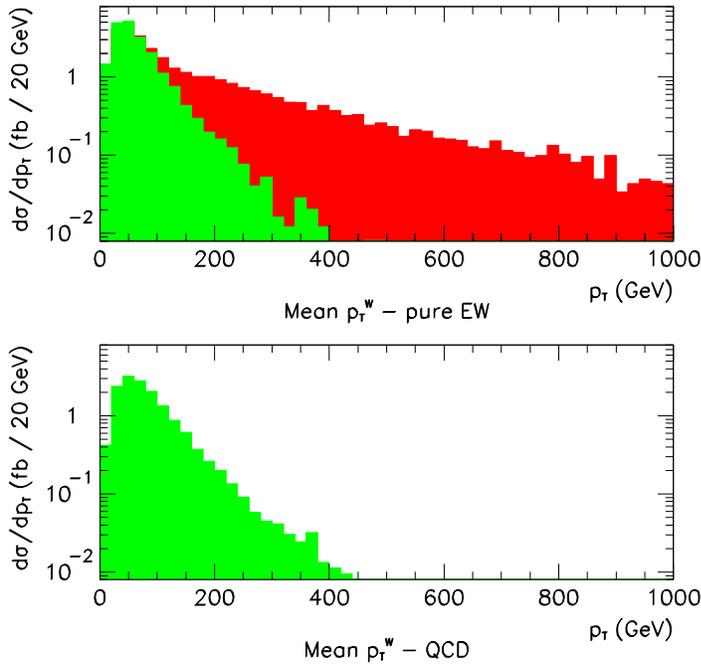,width=0.65\linewidth}
\end{center}
\caption{Transverse momentum distributions of the $W$ coming from pure
  EW production (top) and from QCD production (bottom).  Electroweak
  signal, in the sense of Eq.~(\ref{eq:sigdef}), is shown in red (dark
  grey).  Shown in green (light grey) are contributions to the
  background.}
\label{fig:qcd}
\end{figure}
Moreover, as shown in Fig.~\ref{fig:qcd}, QCD events populate mostly a
different region of kinematic phase space than the electroweak signal
events (mostly apparent in the respective distributions of the $W$
transverse momenta), so interference terms described by the product
$[a(p_T^W,\mathrm{higgsless})_{LL} - a(p_T^W,m_h=120)_{LL}] \cdot
b^{\star}(p_T^W)_{LL}$ also remain negligibly small compared to the
pure EW contribution.  This conclusion is valid as much for $W^+W^+$
as for $W^+W^-$, even though in the latter case QCD processes dominate
the total production rates.

For background that includes also $WW$ bosons with $ij=LT, \,TT$
polarizations, the effect of QCD processes is much more important
since $|b_{ij}|^2$ term is present, and the coherent sum
$a_{ij}\alpha$ does not overwhelm $b_{ij}\alpha_s$.  In fact, QCD
contributions to the total $jjW^+W^-$ production dominate the total
cross section by roughly an order of magnitude.  In particular,
processes involving quark-gluon and gluon-gluon interactions, absent
in the same-sign $WW$ channel, here amount to as much as 60\% of the
total $W^+W^-$ production cross section.  For the same-sign $WW$
production, only quark-quark interactions can contribute and QCD
effects, in form of diagrams involving gluon exchange, amount to
roughly 50\% of the total cross section.  Most of these events
originate from soft parton-parton collisions (soft partons dominate in
the proton PDFs), give very soft kinematics of the accompanying parton
jets and can be rejected by appropriate cuts on their rapidity.

\begin{figure}[htbp]
  \begin{tabular}{ll}
    \epsfig{file=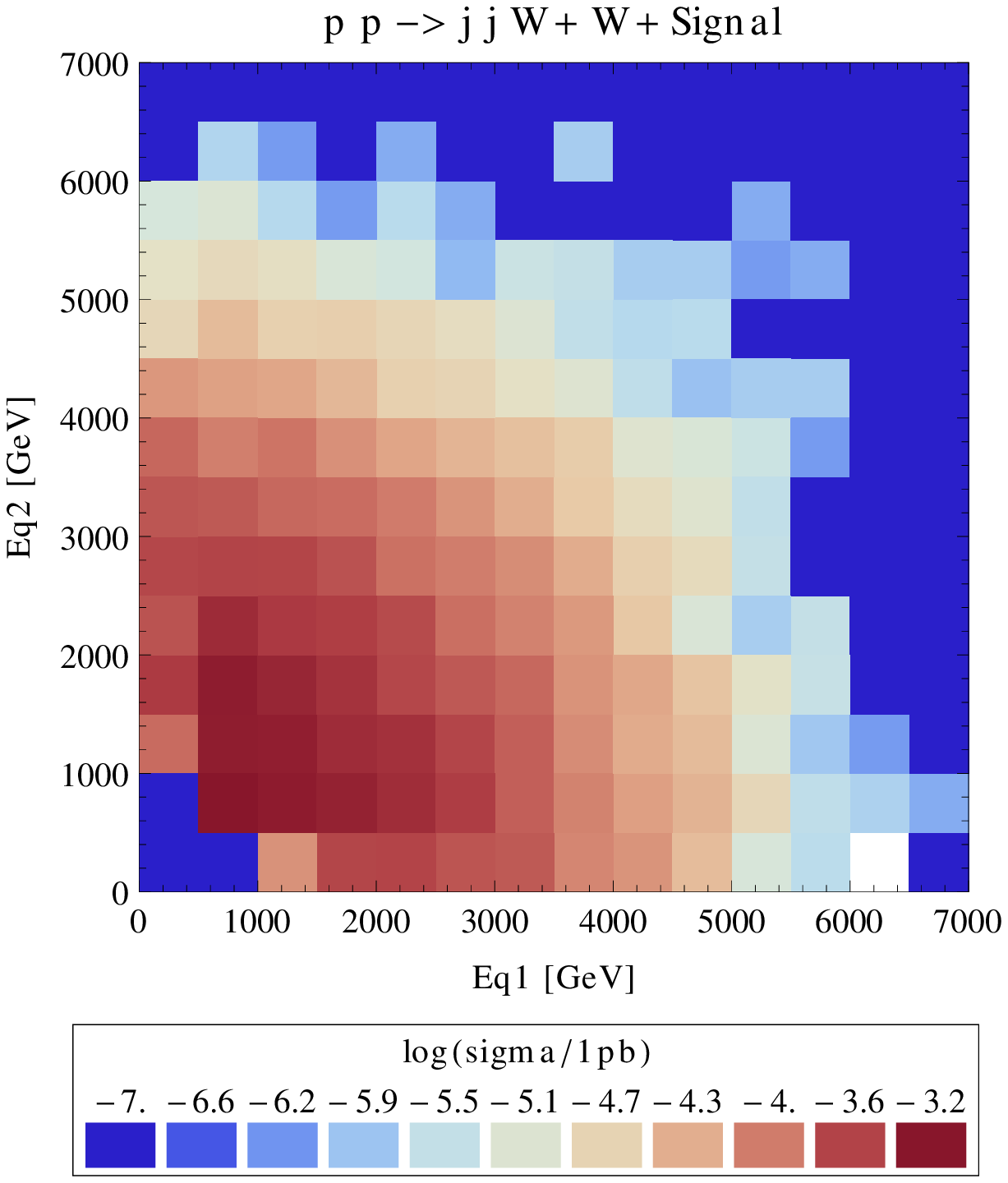,width=0.49\textwidth} &
    \epsfig{file=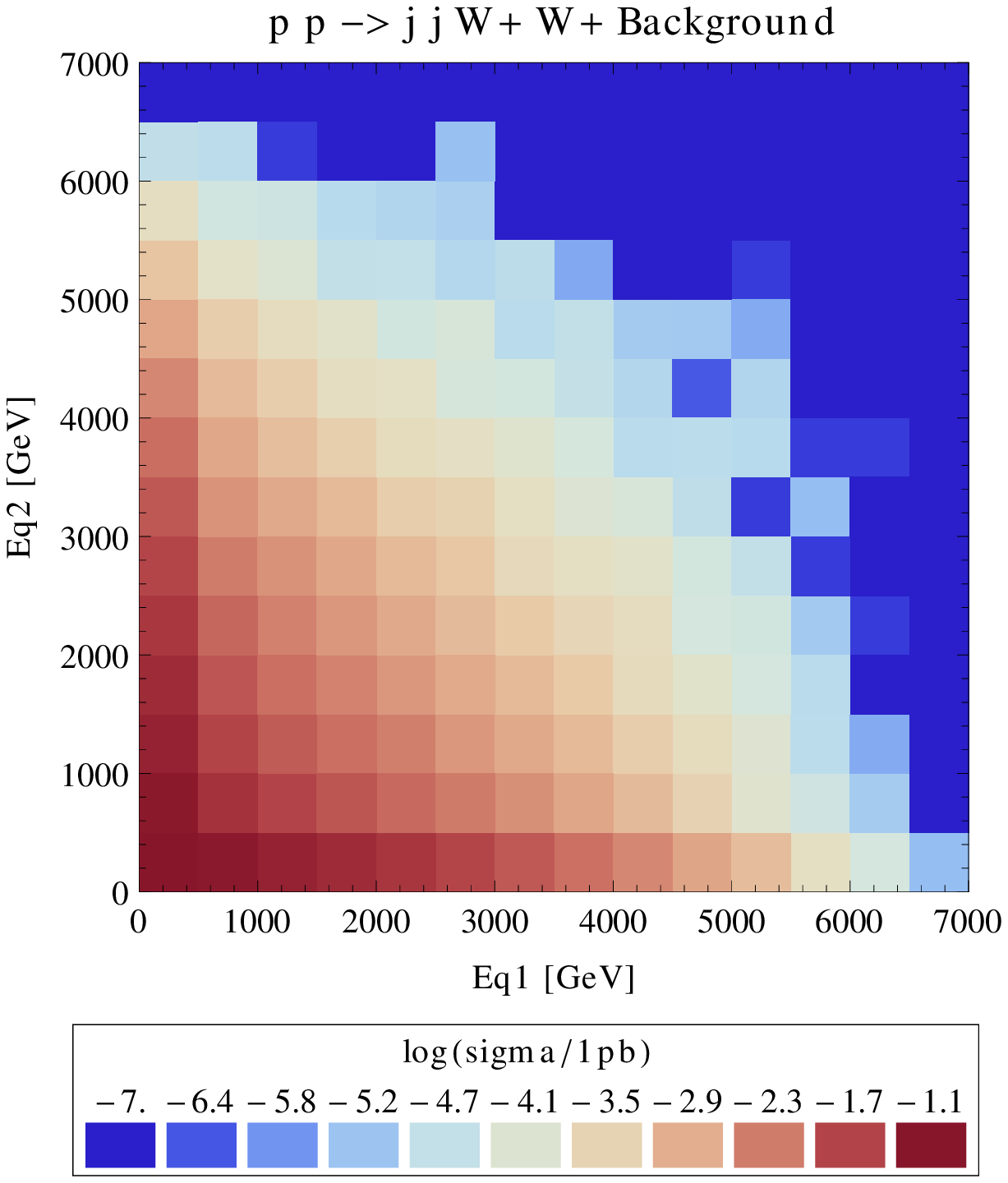,width=0.49\textwidth} \\
    \epsfig{file=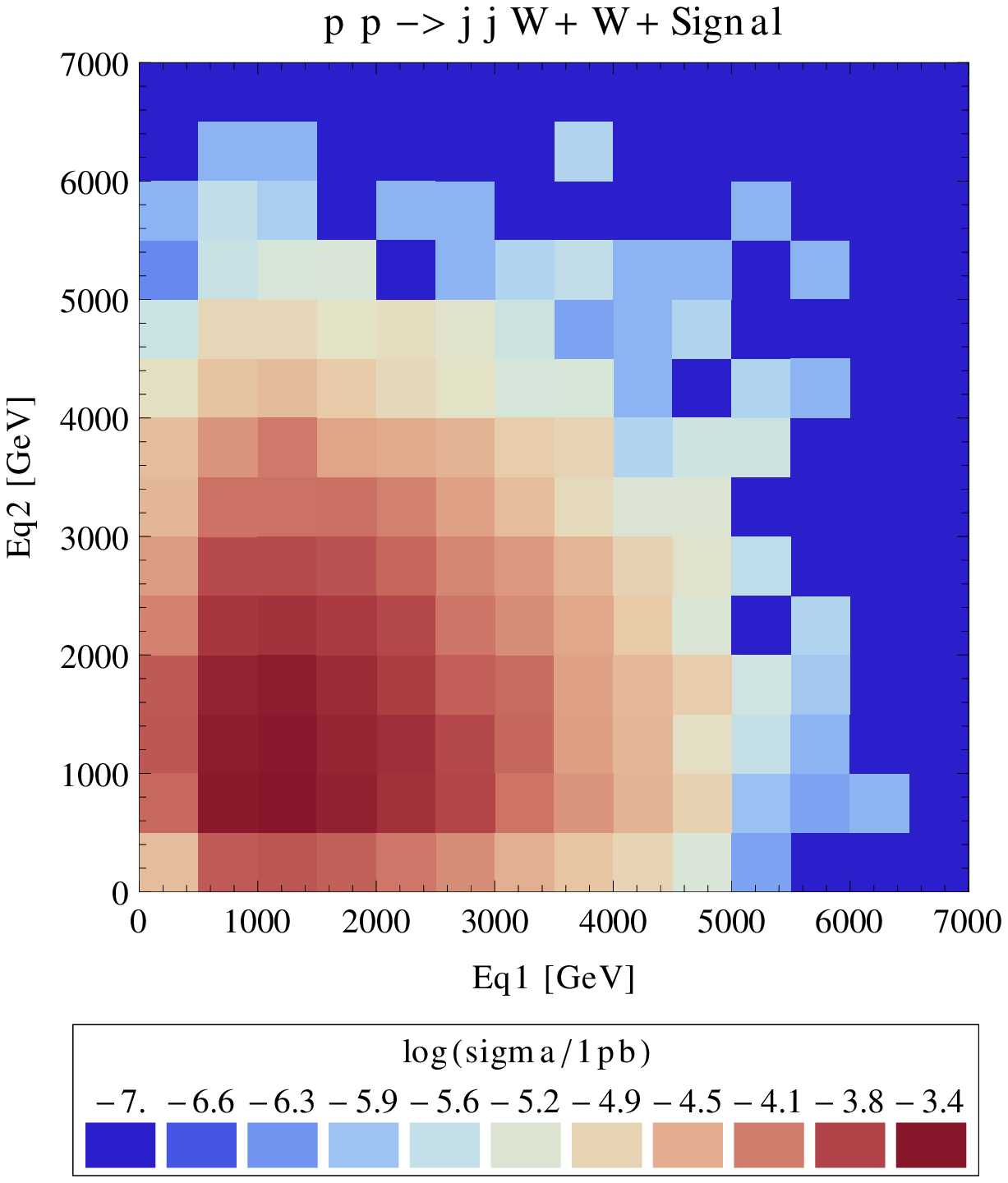,width=0.49\textwidth} &
    \epsfig{file=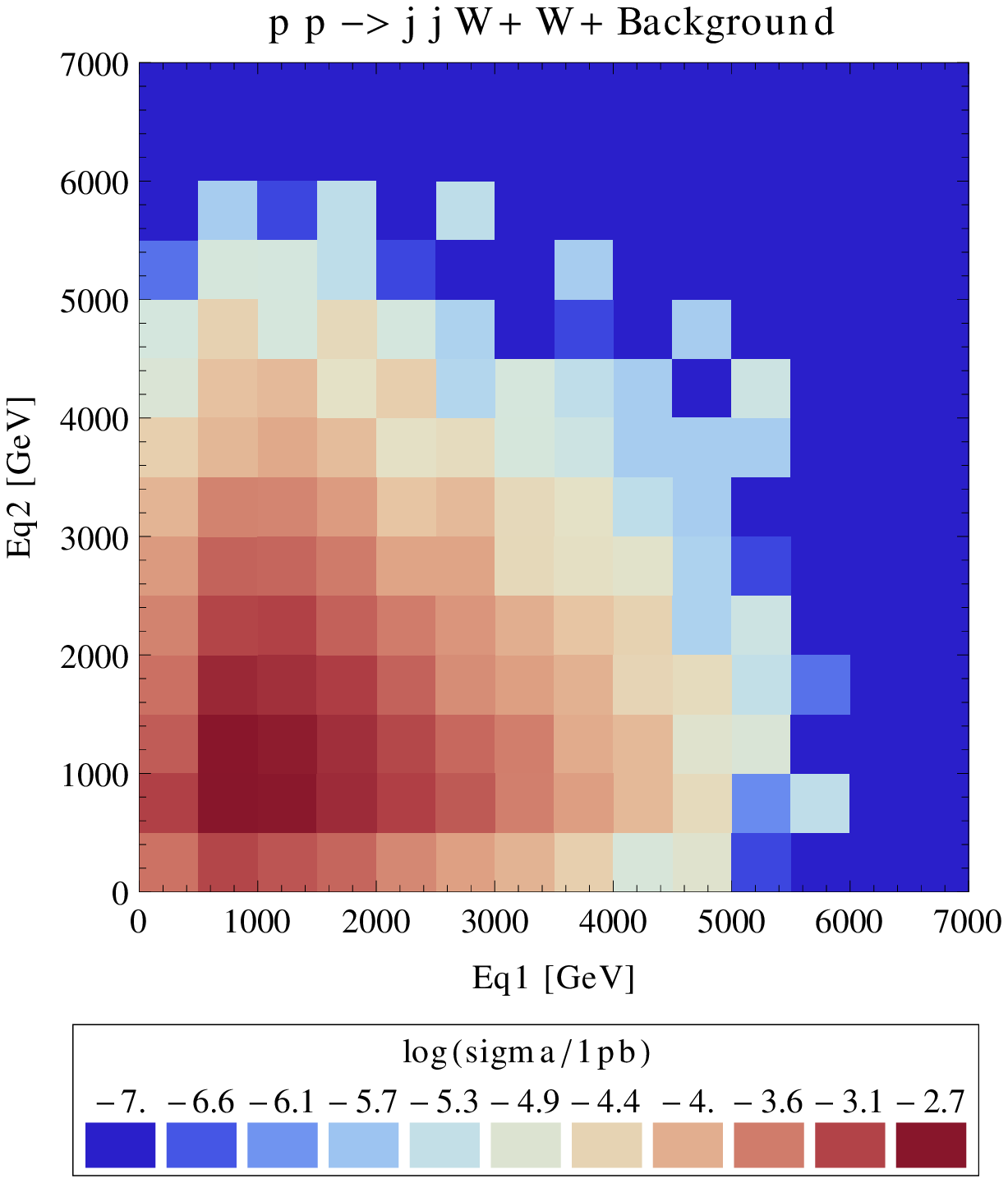,width=0.49\textwidth} \\
\end{tabular}
\caption{Cross section for the $pp \rightarrow jjW^+W^+$ process at 14
  TeV as a function of incident quark energies.  Left column: signal
  as defined in Eq.~(\ref{eq:sigdef}), right column: background
  (Eq.~(\ref{eq:bckgdef})), upper row: without cuts, lower row: after
  ``basic" cuts of Eqs.~(\ref{cut:etaj1},\ref{cut:etaw}).}
\label{fig:eff}
\end{figure}

Conventional selection criteria for $WW$ scattering processes include
the requirement of two tag jets in the forward region and in opposite
directions.  We will quantify this requirement as
\bea
&&2 < |\eta_j| <5 \mathrm{~~ and ~~} \eta_{j1}\cdot\eta_{j2}<0\,.
\label{cut:etaj1}
\eea 
where the $\eta_{j}$ denote pseudorapidities of the parton-level jets.
Detector acceptance imposes stringent limits on the allowed lepton
kinematics to be considered, which at the level of undecayed $W$ can be
approximated as a cut on the $W$ pseudorapidity,
\bea
|\eta_W| < 2\,.
\label{cut:etaw}
\eea 
The effect of these basic cuts is twofold.  As illustrated in
Fig.~\ref{fig:eff}, they significantly reduce the irreducible
electroweak background coming from low energy parton-parton
collisions, that is in the region of small transverse momenta of the
parton-level jets.  They also reduce the QCD background, to the effect
of same-sign $WW$ production being dominated by pure electroweak
production.

Ultimately, a full calculation of the same-sign $WW$ production with
all the proper EW/QCD interference terms included ends up in
background rates that are larger by less than 10\% compared to a pure
EW calculation after all selection criteria are applied and indeed,
many previous studies neglected this contribution altogether.
However, the residual QCD contribution to the background in the
$W^+W^-$ channel is still sizable, including that from quark-gluon and
gluon-gluon interactions and from other processes which change the
overall kinematic characteristic of the background for $W^+W^-$.

In all our MADGRAPH calculations we use a fixed factorization and
renormalization scale of 91.188 GeV.  Our QCD cross sections for the
process $pp \rightarrow jj W^+W^-$ are in agreement to $\sim$10\% with
the respective numbers from Table 7 of Ref.~\cite{zeppenfeld} once
their ``inclusive" cuts are applied, which may well be taken as the
intrinsic uncertainty of our QCD background calculations.  It should
be noted however that unlike the authors of Ref.~\cite{zeppenfeld} we
compute QCD backgrounds together with EW backgrounds, with all due
interference terms in place.

\begin{figure}[htbp]
\begin{tabular}{ll}
  \epsfig{file=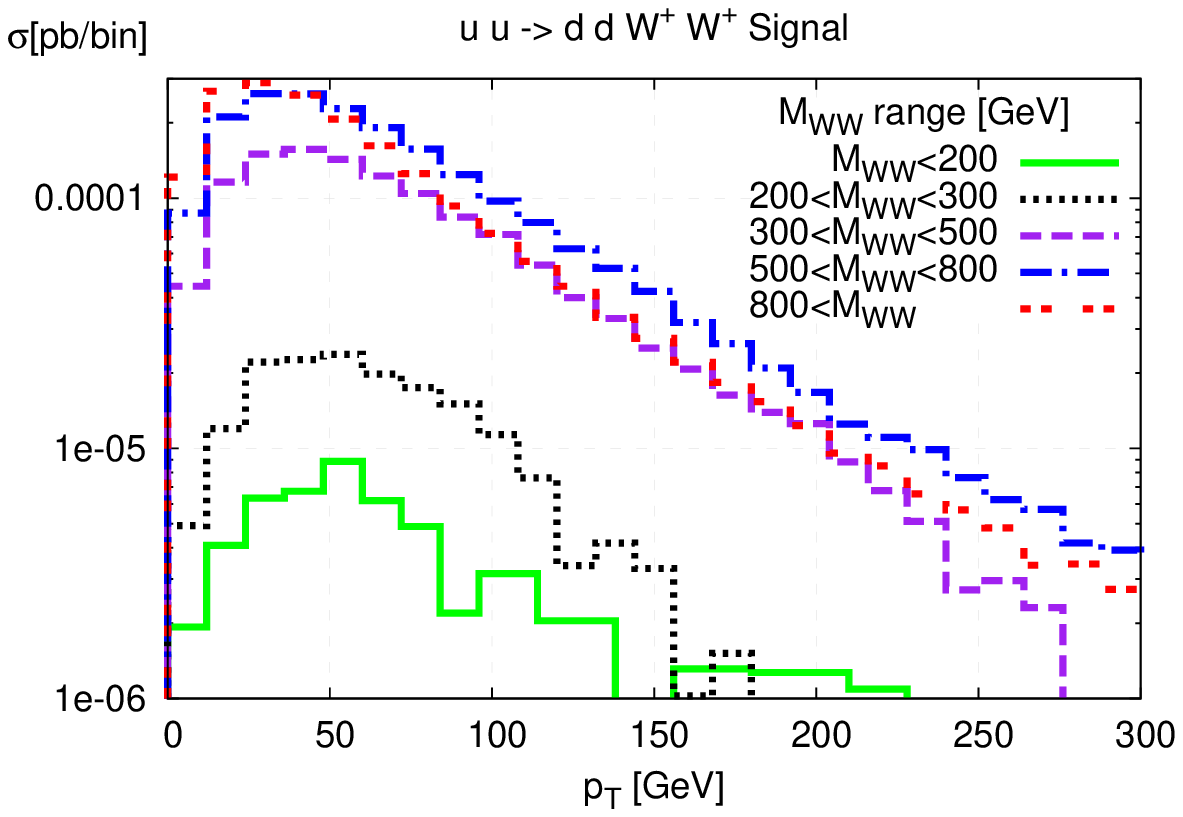,width=0.49\textwidth,height=0.4\textwidth}&
  \epsfig{file=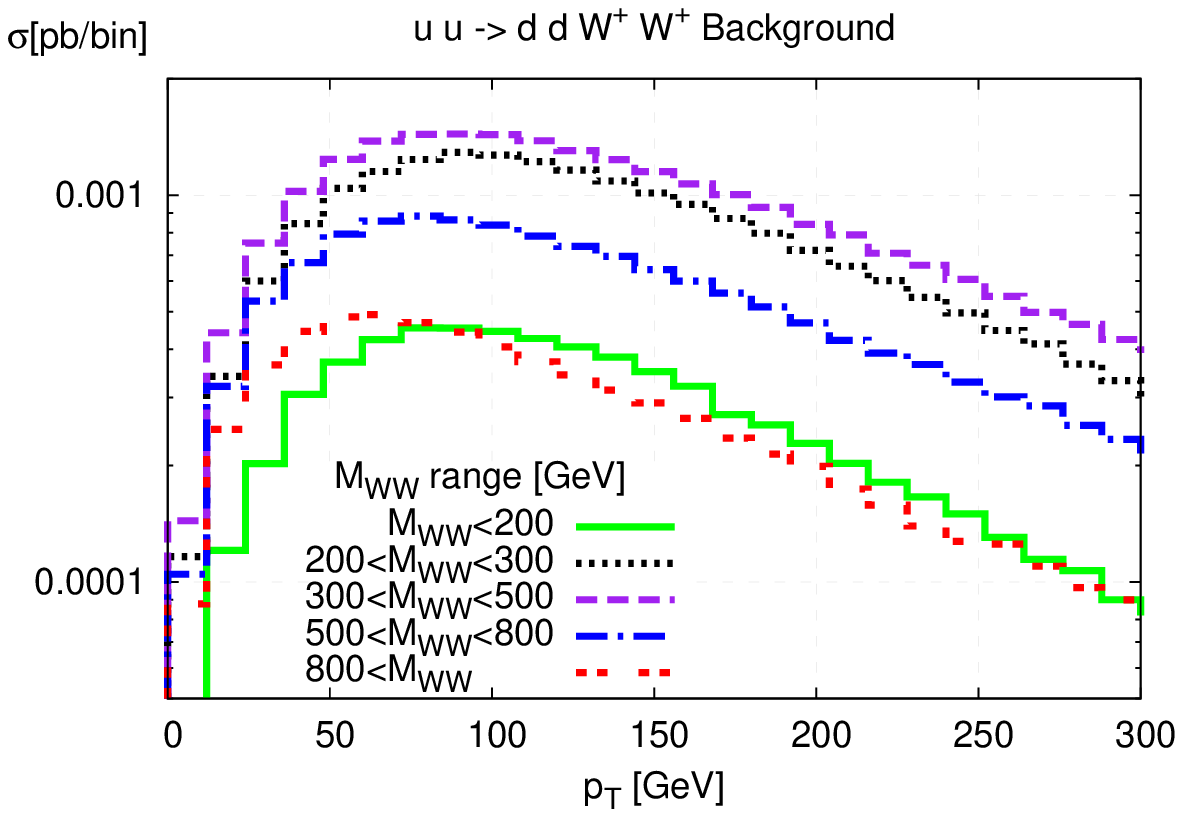,width=0.49\textwidth,height=0.4\textwidth}\\[-2mm]
a) & b)
\end{tabular}
\caption{$p_T$ distributions of parton-level jets in symmetric
  quark-quark collisions $uu \rightarrow jjW^+W^+$ with initial quarks
  of 1 TeV energy: (a) signal, (b) background.  Results from a
  MADGRAPH calculation.}
\label{fig:uu>ddw+w+(ptj,Mww)}
\end{figure}

From Fig.~\ref{fig:eff} we see that basic cuts on the pseudorapidities
of parton-level jets and the outgoing $W$'s in the same-sign channel
case produce a sample of events with little contamination from low
energy quark-quark collisions.  For the background this also means
little QCD contribution left.  At the quark level, the process that
dominates same-sign $WW$ production at the LHC is an approximately
symmetric quark-quark interaction, with its center-of-mass energy
typically peaking at $\sim$$\sqrt{s}/7$, associated with a $W$
emission from each quark line.  Thus the basic features of $pp
\rightarrow jj W^+W^+$ at 14 TeV are qualitatively similar to those of
a pure quark level process $uu\rightarrow dd W^+W^+$ at an energy of
about 2 TeV.  Analysis of this much simpler process is helpful in
deriving effective criteria to separate the $W_LW_L$ scattering signal
from the background that remains after the ``basic" cuts and includes
scattering processes with transverse $W$'s.

\begin{figure}[htbp]
\begin{tabular}{ll}
  \epsfig{file=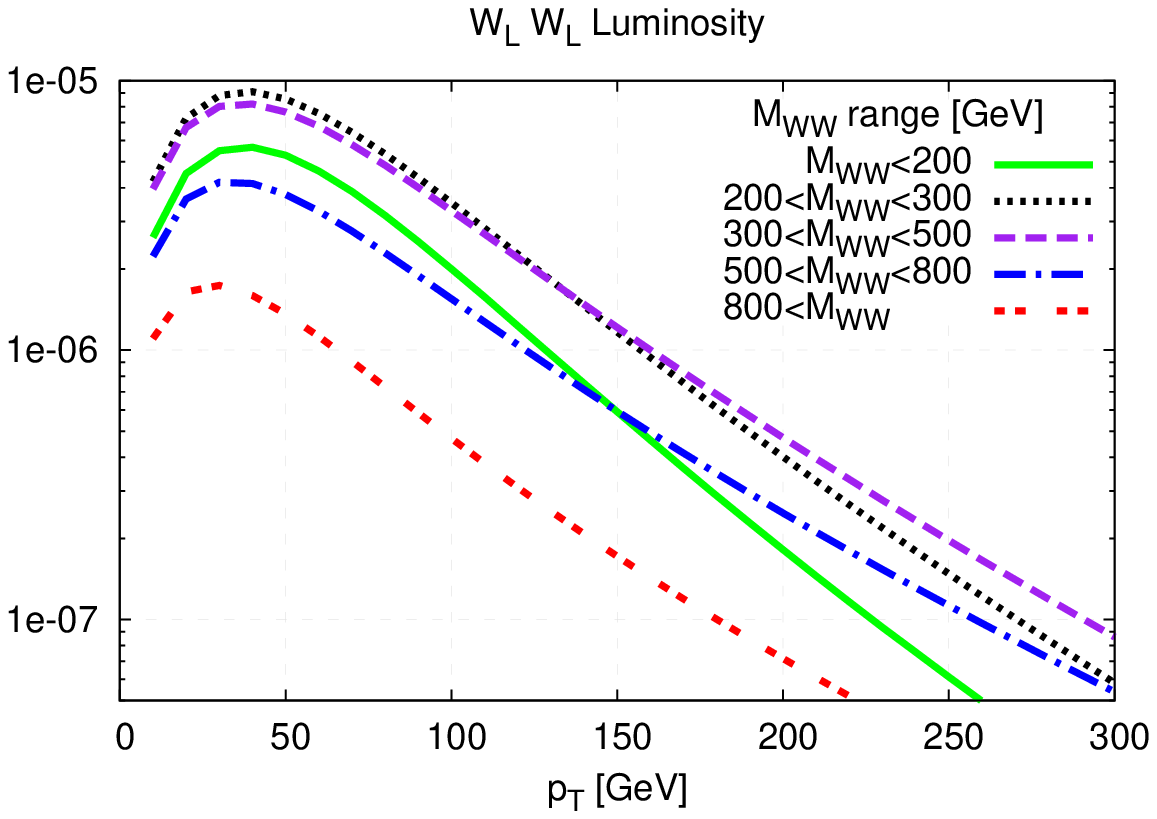,width=0.49\textwidth,height=0.4\textwidth}&
  \epsfig{file=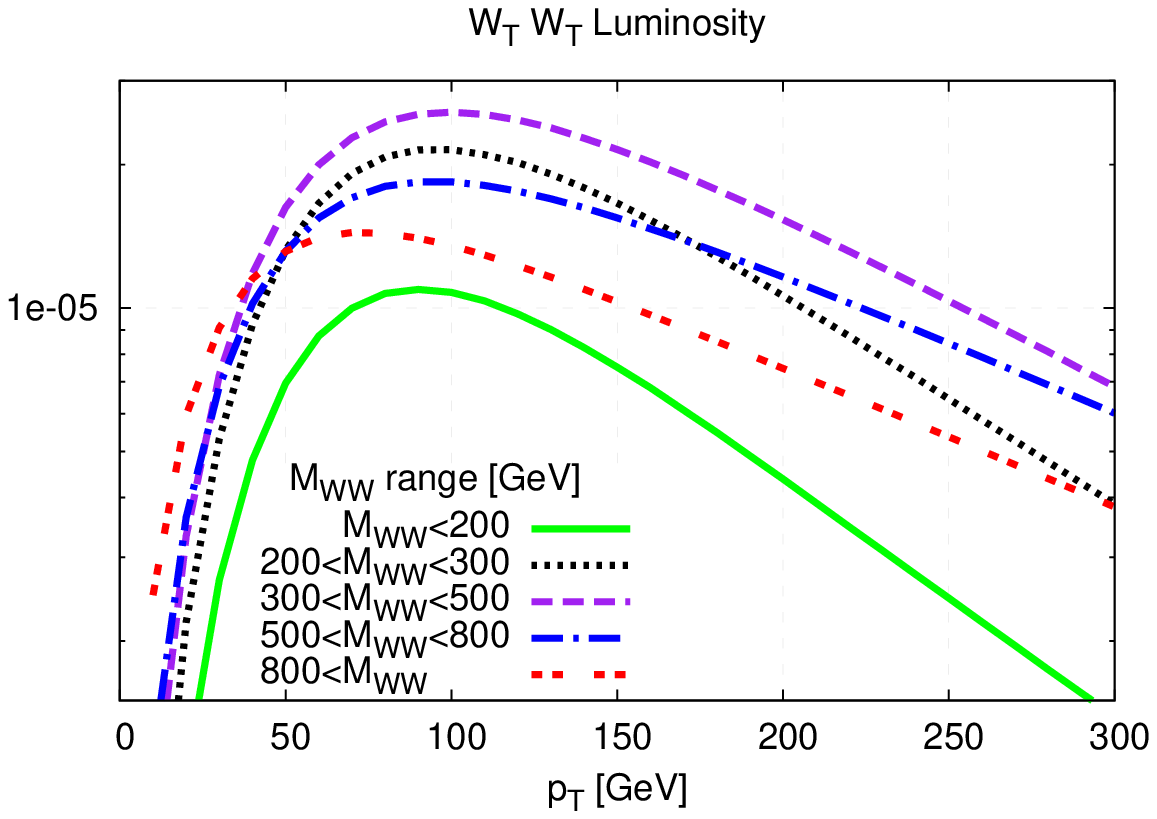,width=0.49\textwidth,height=0.4\textwidth}\\[-2mm]
  a) & b) \\
\end{tabular}
\caption{$p_T$ distributions of parton-level jets associated with $W$
  emission in symmetric quark-quark collisions with initial quarks of
  1 TeV energy: (a) for longitudinal $W$'s, (b) for transverse $W$'s.
  Result (up to an overall normalization factor) of an EWA calculation 
  of the emission probability of two $W$'s, each from a different quark
  line, for given $W$ helicities and $WW$ invariant mass ranges.}
\label{fig:L(ptj,Mww)}
\end{figure}

\begin{figure}[htbp]
\begin{tabular}{ll}
  \epsfig{file=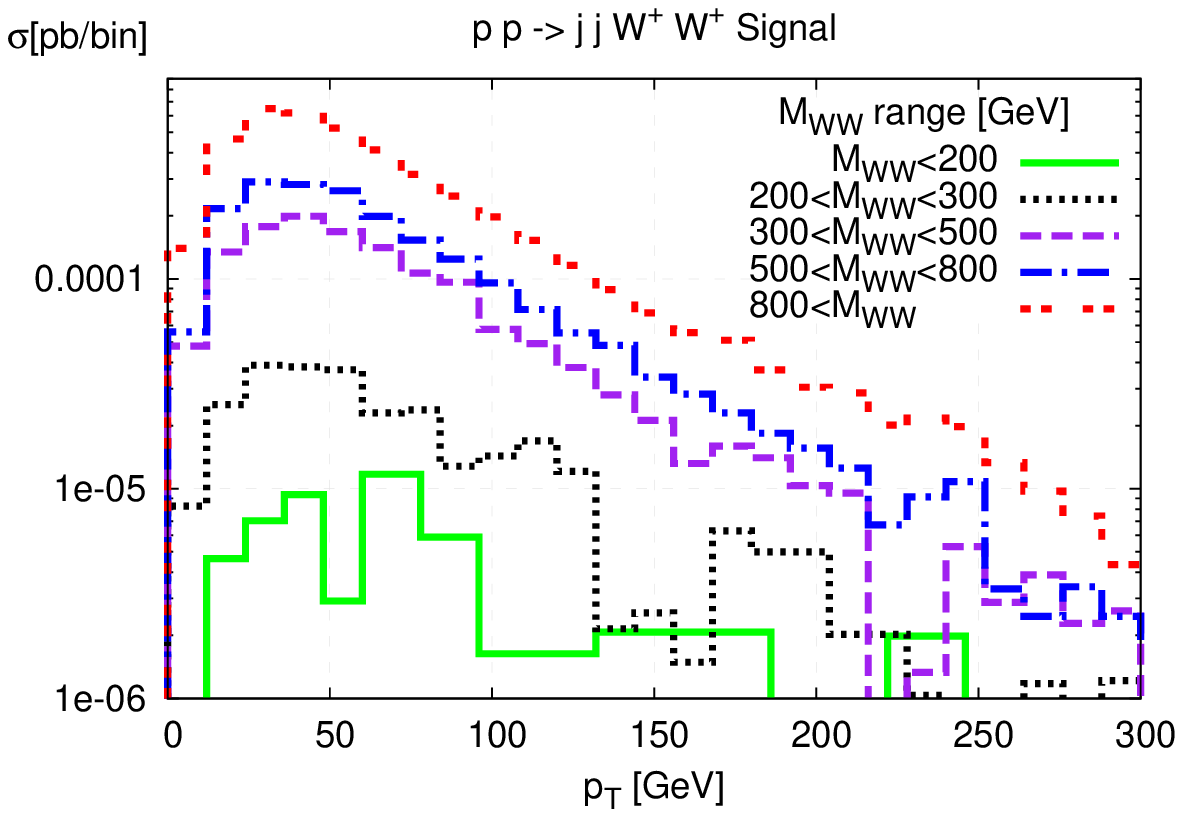,width=0.49\textwidth,height=0.4\textwidth} &
  \epsfig{file=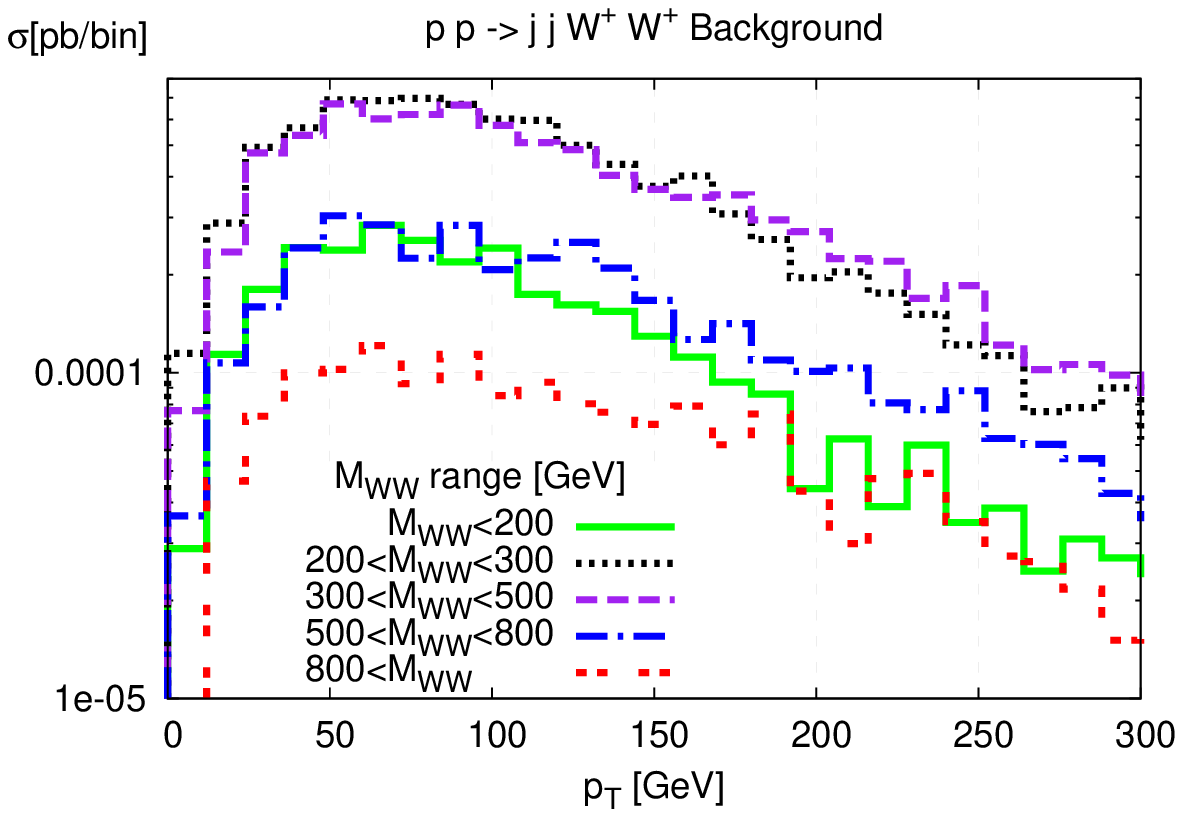,width=0.49\textwidth,height=0.4\textwidth}\\[-2mm]
  a) & b)
\end{tabular}
\caption{$p_T$ distributions of jets in a proton-proton collision $pp
  \rightarrow jjW^+W^+$ with each proton of 7 TeV energy: (a) signal,
  (b) background, after applying the basic cuts described in the text.
  Results of a MADGRAPH calculation.}
\label{fig:pp>jjw+w+(ptj,Mww) cuts}
\end{figure}

The characteristic difference in the kinematics of the LL and TT+TL
final states in the quark-level process are illustrated in
Fig.\,\ref{fig:uu>ddw+w+(ptj,Mww)} where the results of a MADGRAPH
simulation of the $uu \rightarrow ddW^+W^+$ process at 2 TeV are
shown.  The longitudinally polarized $W$ tends to be emitted at a
smaller angle (hence smaller $p_T$) with respect to the incoming quark
direction than the transversely polarized $W$~\cite{dawson, bagger}.  As a
consequence, the final quark accompanying the longitudinal $W$ is more
forward than the one associated with the transverse $W$.  This effect
is more pronounced the larger the invariant mass of the $WW$ pair.
The $p_T$ distributions of quarks associated with $W_L$ emission
become narrower as $M_{WW}$ increases and the peak of the
distributions gradually moves to lower values.  No such trend is
visible for the quarks associated with $W_T$ emission, except for
$M_{WW}$ larger than 800 GeV, where the effects of overall energy and
momentum conservation become significant.

Those qualitative observations suggest that the higher the invariant
mass of the $WW$ pair the easier becomes the isolation of the
longitudinal $WW$ signal from the transverse $WW$ background.  Tagging
of forward jets at a fixed large value of $M_{WW}$ is the ideal
technique to be used.

\begin{figure}[htbp]
\begin{tabular}{ll}
  \epsfig{file=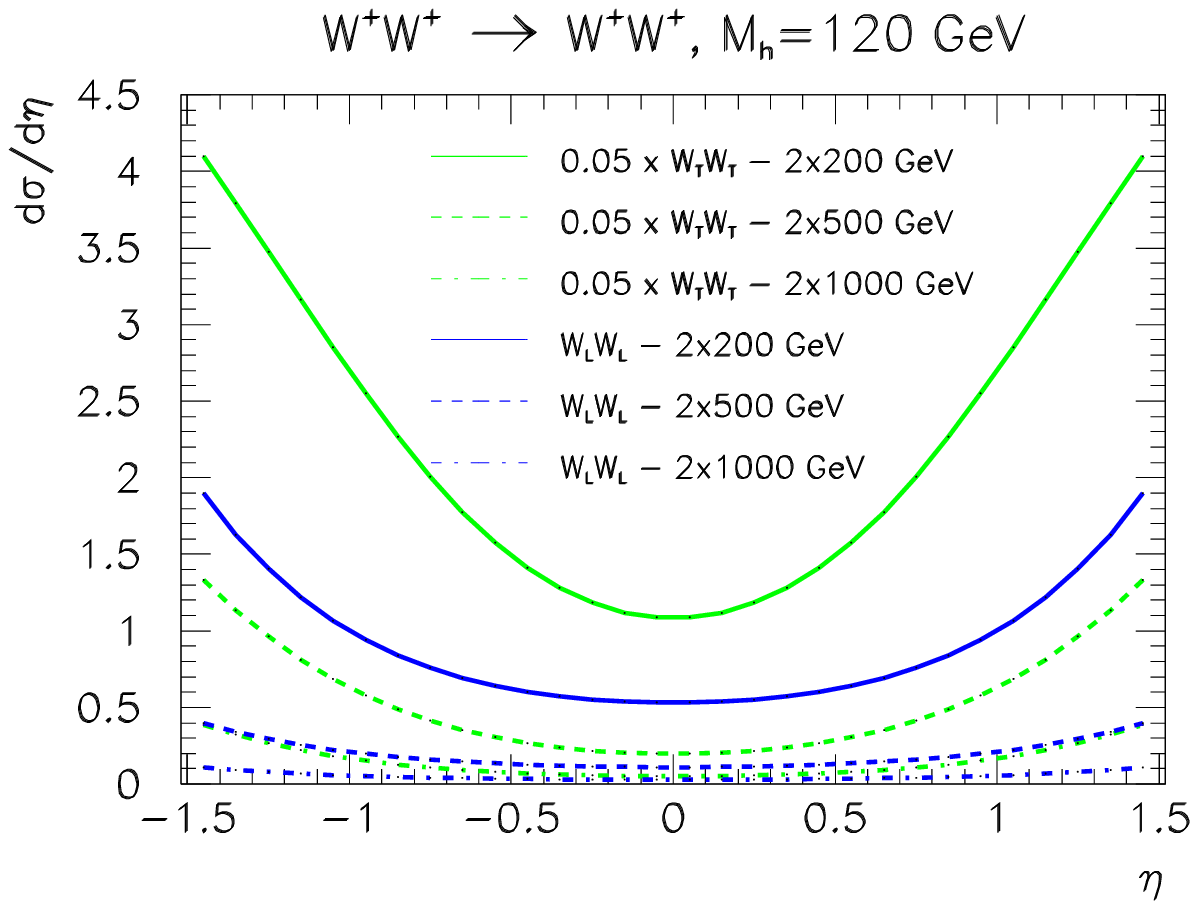,width=0.49\linewidth,height=0.45\textwidth} &
  \epsfig{file=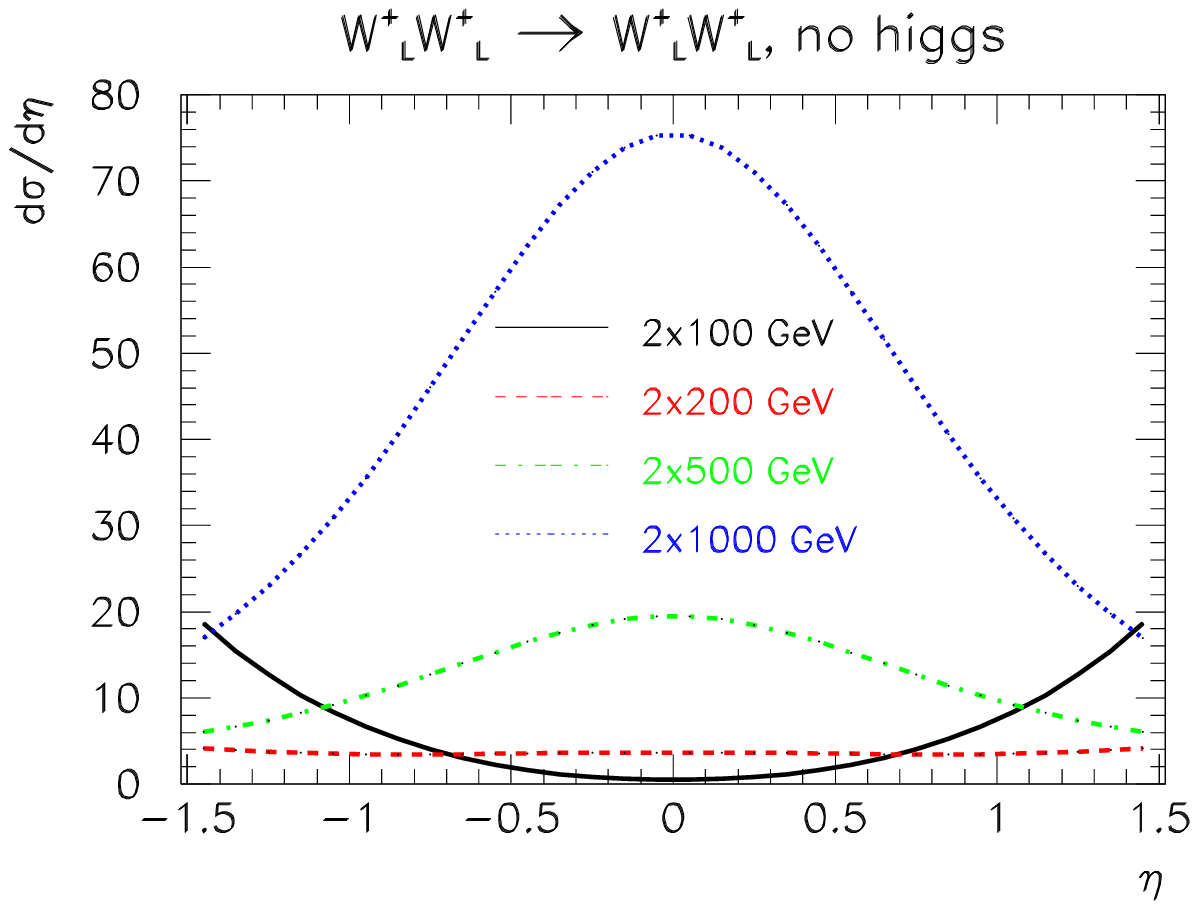,width=0.49\linewidth,height=0.45\textwidth} \\[-2mm]
  a) & b) 
\end{tabular}
\caption{Angular distributions of scattered $W$'s with respect to the
  incoming $W$ direction, in a $W^+W^+$ scattering process at
  different center of mass energies: (a) in the presence of a 120 GeV
  Higgs boson, (b) in the no-Higgs case.}
\label{fig:dsigmadeta}
\end{figure}

It is interesting to note that all these features are qualitatively
reproduced within the framework of the Effective W Approximation,
where we can identify the signal and the background with $W_LW_L$ and
$W_TW_T$, respectively (Fig.~\ref{fig:L(ptj,Mww)}).

In $pp$ collisions, the above regularities are at first completely
masked by the overwhelming low energy and highly asymmetric
quark-quark collisions as well as QCD effects, but become visible
again after the basic cuts discussed earlier.  This is shown in
Fig.\,\ref{fig:pp>jjw+w+(ptj,Mww) cuts}.

The bounds on $\eta_W$ (here mimicking cuts on the pseudorapidities of
$W$ decay products that are necessary to meet detector acceptance
criteria) are instrumental in further reduction of the TT+TL
background.  As shown in Fig.\,\ref{fig:dsigmadeta}, in an ideal
process $WW \rightarrow WW$, with two beams of real $W$'s colliding at
a fixed energy $M_{WW}$, the angular distributions of the scattered
longitudinal and transverse $W$'s (measured in the $WW$ center of mass
and with respect to the incoming $W$ direction) become more different
with the increase of $M_{WW}$.  The cross sections for $WW$ scattering
with polarization flip, e.g.~$W_TW_T\to W_LW_L$, are typically three
orders of magnitude smaller than the polarization-conserving ones and
cannot influence the overall angular characteristics.

As discussed above, jet $p_T$ distributions depend on the emitted $W$
polarization and in particular $W_T$ rejection can be improved by
requiring smallness of the transverse momenta of both jets.  Any
higgsless or non-SM Higgs scenario will modify the angular
distributions of outgoing longitudinal $W$'s by making them more
central with respect to the incoming $W_L$ direction.  Requirement of
a large vector boson $p_T$ measured in the lab frame combines the
effects of a large $W$ scattering angle and small $W$ emission angle
from the parent quark line, both of which favor longitudinal over
transverse $W$'s.  In search for the most efficient selection
criterion that would combine low jet $p_T$ and large vector boson
$p_T$, we notice that in the small $|\eta_W|$ region the $M_{WW}$
value strongly correlates with the product of the two transverse
momenta, $p_T^{W1} \cdot p_T^{W2}$.  In the leptonic $W$ decay
channels, where we do not know the exact kinematics of the two
decaying $W$'s, the $p_T$ values of the leptons and in particular
their product are the best practical measures of $M_{WW}$.
Consequently, a large value of the product $p_T^{l1} \cdot p_T^{l2}$
is an approximate experimental signature of the kinematic region which
is most sensitive to the actual mechanism of electroweak symmetry
breaking and offers the best capabilities to separate longitudinal
from transverse $WW$ scattering processes in any scenario that
enhances the $W_LW_L$ cross section for large $M_{WW}$.  Correlation
between high $p_T^{l1} \cdot p_T^{l2}$ and high $M_{WW}$ is stronger
if one additionally asks for large ratios of $p_T^l/p_T^j$.  For high
pseudorapidity jets this means removing events with very large jet
energies and little energy left for the $W$.

The above considerations make it clear that a large value of the
following ratio $R_{p_T}$ of the four transverse momenta:
\bea
R_{p_T}= {p_T^{~l_1}\cdot p_T^{~l_2}\over p_T^{~j_1}\cdot p_T^{~j_2}},
\eea
where $l_1$ and $l_2$ denote the two leptons in no particular order
and $j_1$ and $j_2$ denote the two most energetic jets in the event,
is bound to have a large efficiency in isolating hard $W_LW_L$
scattering from the SM background.  Such ratio automatically accounts
for the correlations that are likely to be satisfied by the signal,
but not by the background, and thus work more efficiently than a
collection of uncorrelated cuts on the individual variables.

In the rest of the paper we discuss the impact of using the new
variable as a selection criterion and the resulting perspectives for
the observation of $W_LW_L$ scattering at the LHC.  In particular, in
Section~\ref{sec:sameww} we compare the selection efficiencies with
and without the variable $R_{p_T}$ in the analysis of the same-sign
$WW$ channel and demonstrate that the new criterion can significantly
improve the $S/B$ figures for this process.

The kinematics of signal and background for the opposite-sign channel
differs from that of the same-sign channel in several significant
ways.  Apart from the residual QCD background which softens the
average jet $p_T$ and worsens its separation from the signal, pure EW
background receives additional contributions from $t$-channel
processes in which both the $W^+$ and the $W^-$ originate from the
same parent quark line.  This is another source of softening of the
average jet $p_T$ for jets accompanying $W_T^+W_T^-$ pairs and a class
of processes not covered in the EWA approach.  On top of that, both
signal and background receive contributions from $s$-channel processes
with a $W^+W^-$ pair being produced from a $Z$ or a Higgs boson.
Finally, parton distribution functions also play some role, since
the two valence $u$ quarks of the proton favor $W^+W^+$ production.
All these effects change the overall kinematics of the signal
and, most importantly, of the irreducible background.  In
Section~\ref{sec:oppww} we show that these features indeed reduce the
practical usefulness of the $R_{p_T}$ variable in the analysis of the
opposite-sign channel.

\section{Reducible background}
\label{sec:bckg}

Among the many potential sources of reducible background, inclusive
$t\bar{t}$ production appears to be most difficult to suppress.
Having this background under control is of course most critical in the
$jjW^+W^-$ channel where top decays can directly fake the signal.  It
turns out however that also in the $jjW^+W^+$ channel, due to the huge
initial cross section for $t\bar{t}$ production, even tiny detector
effects cannot be completely disregarded and can contaminate the
signal to measurable amounts, necessary to estimate and subtract.  The
two main mechanisms for a $t\bar{t}$ pair to fake the same-sign $WW$
scattering signal are: lepton sign misidentification and leptonic $B$
decays.

We have developed and implemented two independent methods to estimate
the magnitude of the inclusive top pair production.  The first method
relies on an analysis of two separate samples generated with MADGRAPH:
one of $pp \rightarrow t\bar{t}$ and another of $pp \rightarrow
t\bar{t}q$ (top pair production with an associated light quark).  Both
samples were then processed with PYTHIA to account for the effects of
initial and final state radiation, top decay, hadronization and jet
formation.  Since the respective sets of Feynman diagrams which are
included in the calculation of the two samples are mutually exclusive,
the procedure does not involve any event double counting and the
events can be simply added at the end.  The caveat of this method is
that it does not encompass all possible processes leading to top pair
production with one or two additional jets, since diagrams involving
gluon emission off an internal quark line are neglected here.
However, our cross section calculation for $\sqrt{s} = 7$ TeV gives a
total of 198 pb, which is in fair agreement with already published CMS
and ATLAS data~\cite{topcms}.

In the second method we generate three samples using MADGRAPH: $pp
\rightarrow t\bar{t}$, $pp \rightarrow t\bar{t}j$ and $pp \rightarrow
t\bar{t}jj$, with $j = q, g$, then we likewise use PYTHIA for initial
and final state radiation, top decay, hadronization and jet formation.
To avoid double counting of the events we follow the prescription
described in more detail in Ref.~\cite{zeppenfeld}, which amounts to
selecting from each sample only the events with mutually exclusive
topologies.  In our case, we keep only events in which the number of
$b$ quarks outgoing at a pseudorapidity $2 < |\eta| < 5$ is 2, 1 or 0,
respectively.  The three samples thus represent the respective lowest
order processes for which 0, 1 or 2 tagging jets arise from gluons or
light quarks.  This method is formally more complete and coherent,
even though the total cross section without cuts cannot be deduced and
confrontation with existing experimental data must necessarily rely on
comparisons with results obtained from the former method.  All the
$t\bar{t}$ numbers we will quote in this paper are to be understood as
arising specifically from the latter method and having been
additionally cross checked with the former.

In the determination of $t\bar{t}$ background we will be assuming
throughout this paper an average $b$-tagging efficiency of 50\% for a
single $b$, with a negligible ($\sim1\%$) probability of mistagging a
lighter quark or gluon, in reasonable consistency with recent CMS
reports~\cite{cmsbtag}.

Determination of background arising from leptonic $B$ decays requires
final state radiation switched on and some jet reconstruction
procedure to be adopted.  In this work we used the jet reconstruction
algorithm which is provided by PYTHIA routine PYCELL, with the cone
width set to 0.7.  For consistency we also used exactly the same
procedure for the irreducible background as well as for the signal.
Compared to a pure parton-level analysis, this implies a decrease in
the number of accepted signal events by nearly 10\%.
\begin{figure}[htb] 
\begin{center}
\epsfig{file=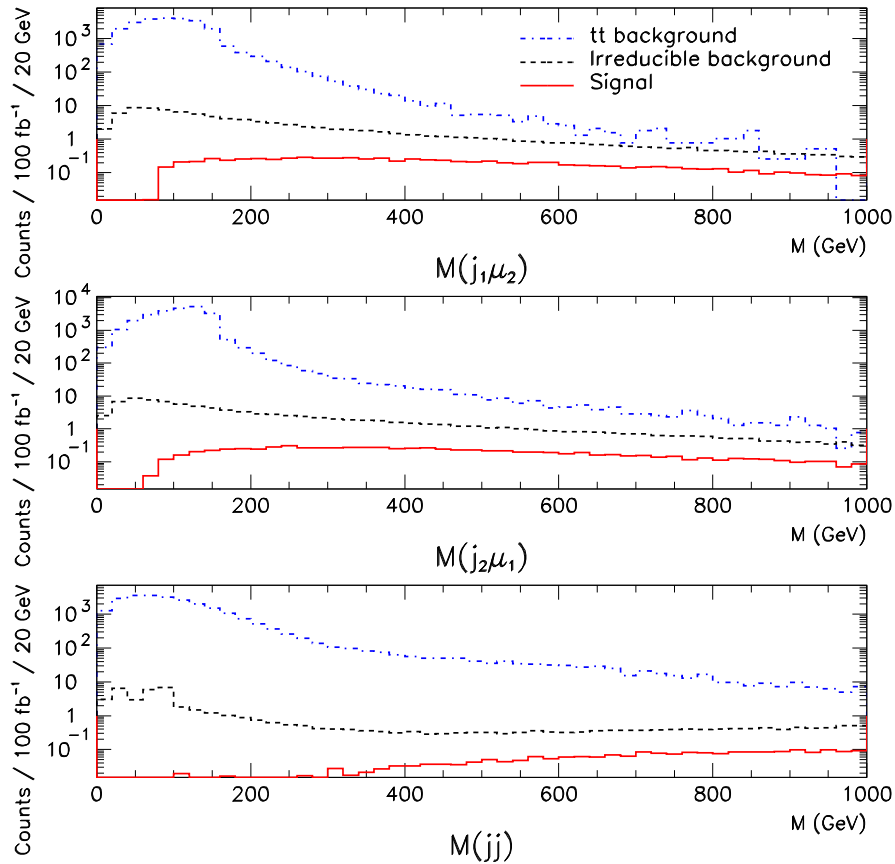,width=0.8\linewidth}
\end{center}
\caption{Kinematic distributions of the $t\bar{t}$ background compared
  to signal and irreducible background in the $pp \rightarrow
  jj\mu^+\mu^+\nu\nu$ process at 14 TeV: Invariant masses of two
  leading jets (bottom) and combinations of jets and muons (top and
  middle).}
\label{fig:signalbackground}
\end{figure}

The primary selection criteria used to suppress the $t\bar{t}$
background consist of stringent cuts on the invariant mass of the two
most energetic jets in the event and combinations of jets with
leptons, see Fig.\,\ref{fig:signalbackground}.  Based on these
distributions we chose to apply cuts on the invariant masses of
combinations $j_1l_2$ and $j_2l_1$ (both jets and leptons are ranked
according to their transverse momenta) at 200 GeV.  Cuts on the other
two jet-lepton combinations are optional, we find however they do not
improve the final figure of merit, defined as $S/\sqrt{S+B}$, and
hence we drop them.  We also apply a cut on the jet-jet invariant mass
whose actual value is subject to individual optimization in each
analysis.

Furthermore, in order to estimate some reasonable event yields in the
analysis of the same-sign $WW$ channel, we will assume a realistic
efficiency of muon sign matching of 99.5\% for $p_T \sim 300$
GeV~\cite{cmsmuonsign}.  For electrons at $|\eta|<2$, a 99\% sign
matching capability with an overall electron reconstruction efficiency
of 95\% seems attainable in CMS~\cite{cmselereco} and these values
will be assumed in the rest of the paper.

We have not studied any possible additional sources of reducible
background in this work since they have been shown less relevant in
both the $jjW^+W^+$ and $jjW^+W^-$ channels (e.g.~\cite{other_bckg})
and partly because most of these backgrounds involve various detector
effects which cannot be studied without a proper detector simulation.
In particular, one should be mindful of the potentially harmful
additional reconstruction backgrounds in the electron channel, notably
the ``fake electron" background, but such considerations surpass the
scope of the present work.

\section{The same-sign $WW$ channel}
\label{sec:sameww}

We have generated the following samples: $jjW^{\pm}W^\pm$ in the SM
with a 120 GeV Higgs (accounting for the total irreducible
background), $jjW_L^{\pm}W_L^\pm$ in the SM with a 120 GeV Higgs and
in the higgsless case (for the signal calculation) and several
$t\bar{t}$ samples generated as described in the previous Section.
Generated numbers of events and the respective cross sections,
together with other details, are shown further in
Tables~\ref{tab:results} and~\ref{tab:resultsmm}.  The only generation
level cut was $2 < |\eta_j| < 5$ for both outgoing parton-level jets
(except for the $t\bar{t}$ samples, where no generation level cuts
were applied), to avoid the most background dominated kinematical
region and satisfy detector acceptance.

The conventional approach to decrease the irreducible background
consists of applying a set of uncorrelated leptonic cuts in addition
to jet cuts.  As discussed in Section~\ref{sec:prod+em}, leptons
originating from the decay of hard $WW$ scattering processes tend to
have larger transverse momenta, they are more central, more
back-to-back and have larger invariant masses than those from
background processes.  Finally, the bulk of the background from
$t\bar{t}$ with a $B$ meson decaying into leptons can be efficiently
suppressed by lepton isolation techniques which we will approximate
here with a requirement of no reconstructed jets within a cone of 0.4
centered at each lepton.  We also require no tagged $b$ and no
additional leptons with $p_T > 10$ GeV within $|\eta|<2$.  A complete
set of conventional signal selection criteria can therefore be as
follows (set I):

\begin{itemize}
\item exactly 2 same sign leptons within detector acceptance,
\item 2 tag jets with $2 < |\eta_j| < 5$ and opposite directions,
\item no $b$-tag,
\item $M_{j_1l_2}$, $M_{j_2l_1} >$ 200 GeV,
\item $M_{jj} >$ 400 GeV,
\item $\Delta R_{jl} >$ 0.4,
\item $p_T^{~l_1}$, $p_T^{~l_2} >$ 40 GeV,
\item $|\eta_{l_1}|$, $|\eta_{l_2}| <$ 1.5,
\item $\Delta\phi_{ll} >$ 2.5,
\item $M_{ll} >$ 200 GeV.
\end{itemize}

\begin{figure}[htbp]
  \begin{tabular}{ll}
    \epsfig{file=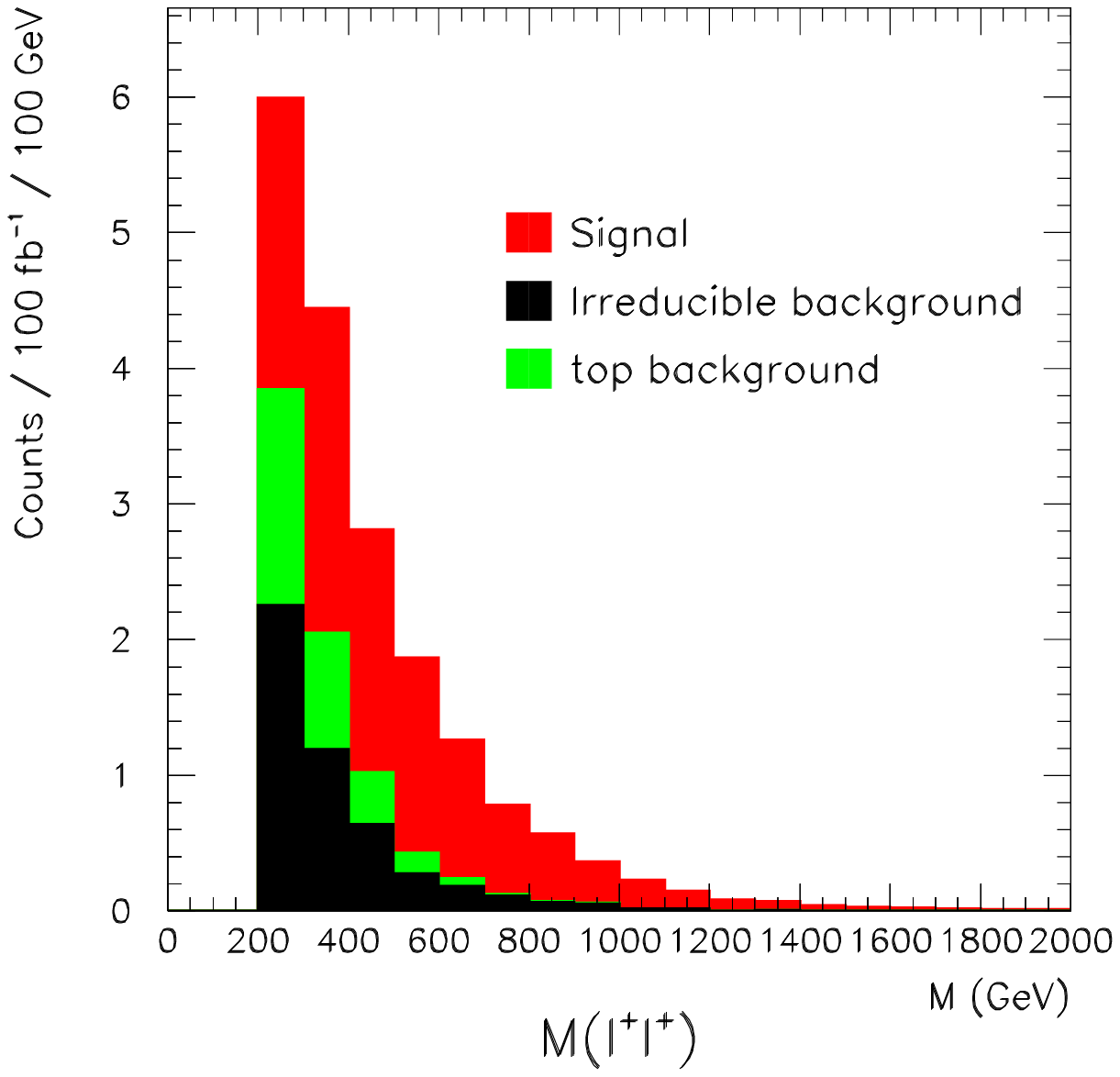,width=0.49\linewidth} &
    \epsfig{file=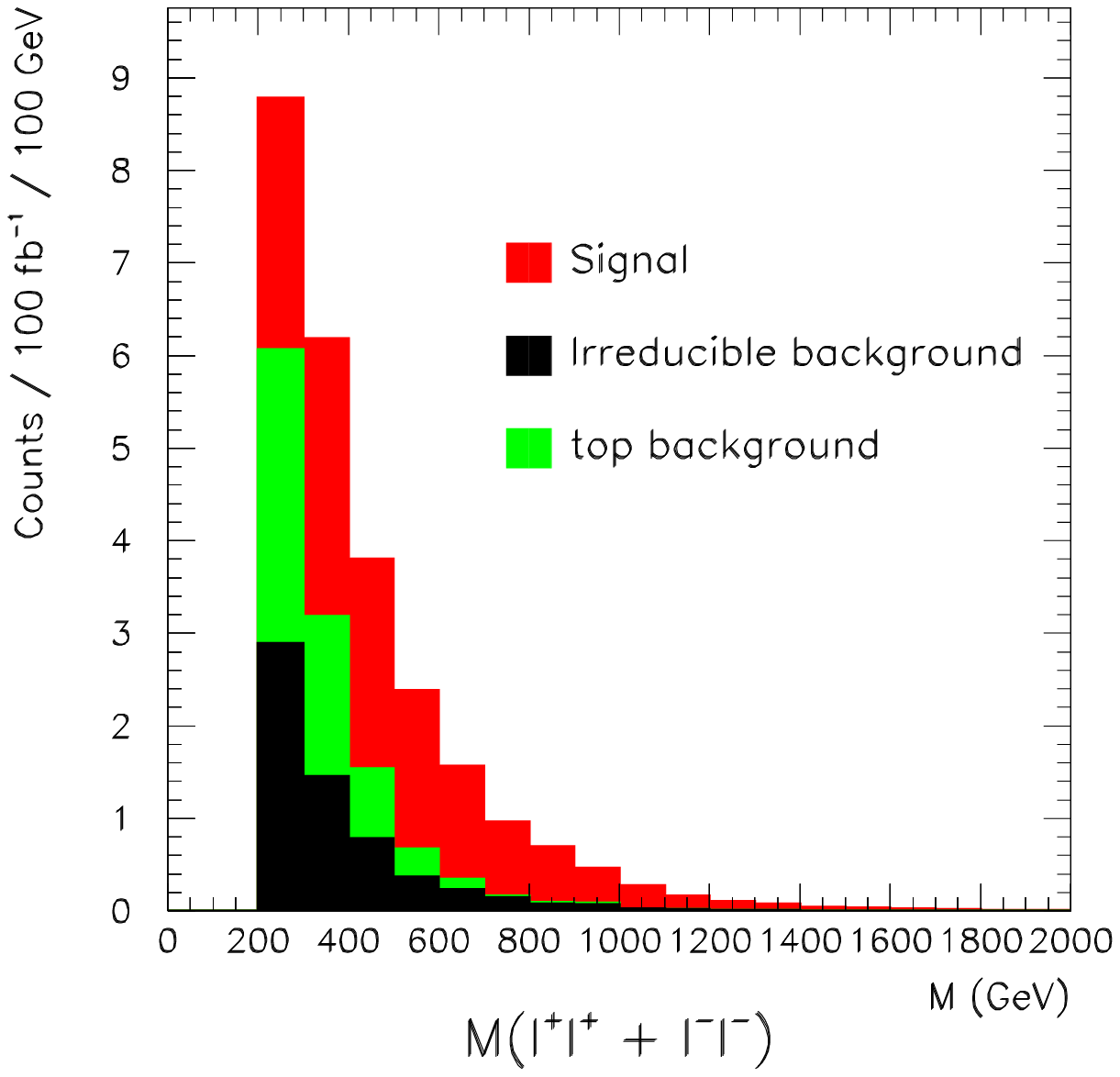,width=0.49\linewidth} \\
    a) & b) 
\end{tabular}
\caption{Invariant mass distributions of two same-sign leptons after
  all the conventional signal selection criteria (set I), normalized
  to 100 $fb^{-1}$: (a) $jjl^+l^+$, (b) $jjl^+l^+$ and $jjl^-l^-$
  added.}
\label{fig:results_old}
\end{figure}

After adding all the same-sign leptonic channels together ($l^+ =
\mu^+, e^+$), we get selection efficiencies and final cross sections
that are given in Table\,\ref{tab:results} and the $l^+l^+$ invariant
mass of the selected events is shown in
Fig.\,\ref{fig:results_old}(a).  Normalized to 100 $fb^{-1}$, we get a
signal to background $S/B \approx 11/7$.  It should be stressed here
that surviving background includes a measurable amount of $t\bar{t}$
and that its dominant source at this point is lepton sign
misidentification.  The inclusion of the $W^-W^-$ channel provides an
additional 25\% to the signal and irreducible background (see
Table\,\ref{tab:resultsmm}), but at the same time the $t\bar{t}$
background doubles, ending up in $S/B \approx 13/11$.  The
corresponding $ll$ invariant mass distribution is shown in
Fig.\,\ref{fig:results_old}(b).  Although our set of cuts is not very fine
tuned, it already allows to produce a result which is roughly
comparable with previous analyses in the same-sign lepton channel, in
particular with the recent Ref.~\cite{ballestrero}.  Several
differences exist between the two analyses (in order of relevance: no
$t\bar{t}$ background included, no unitarity limit applied in the
no-Higgs case, cuts done on parton level variables, different
treatment of QCD background and a different default Higgs mass).
Recalculating under their assumptions, our result translates into $S/B
\approx 17/7$ per 100 $fb^{-1}$, to be compared with their best result
$S/B = 13/6$ \footnote{To derive this number, we take the row
  corresponding to $M_{cut} = 400$ GeV from Table 4 of
  Ref.~\cite{ballestrero}, treat the ``$M_H = 200$ GeV" cross section
  as the irreducible background and subtract it from the ``no Higgs"
  cross section to get the signal, then normalize the result to 100$
  fb^{-1}$.}  (for a discussion of the differences, see
Section~\ref{sec:summary}).

\begin{figure}[htbp]
\begin{tabular}{ll} 
  \epsfig{file=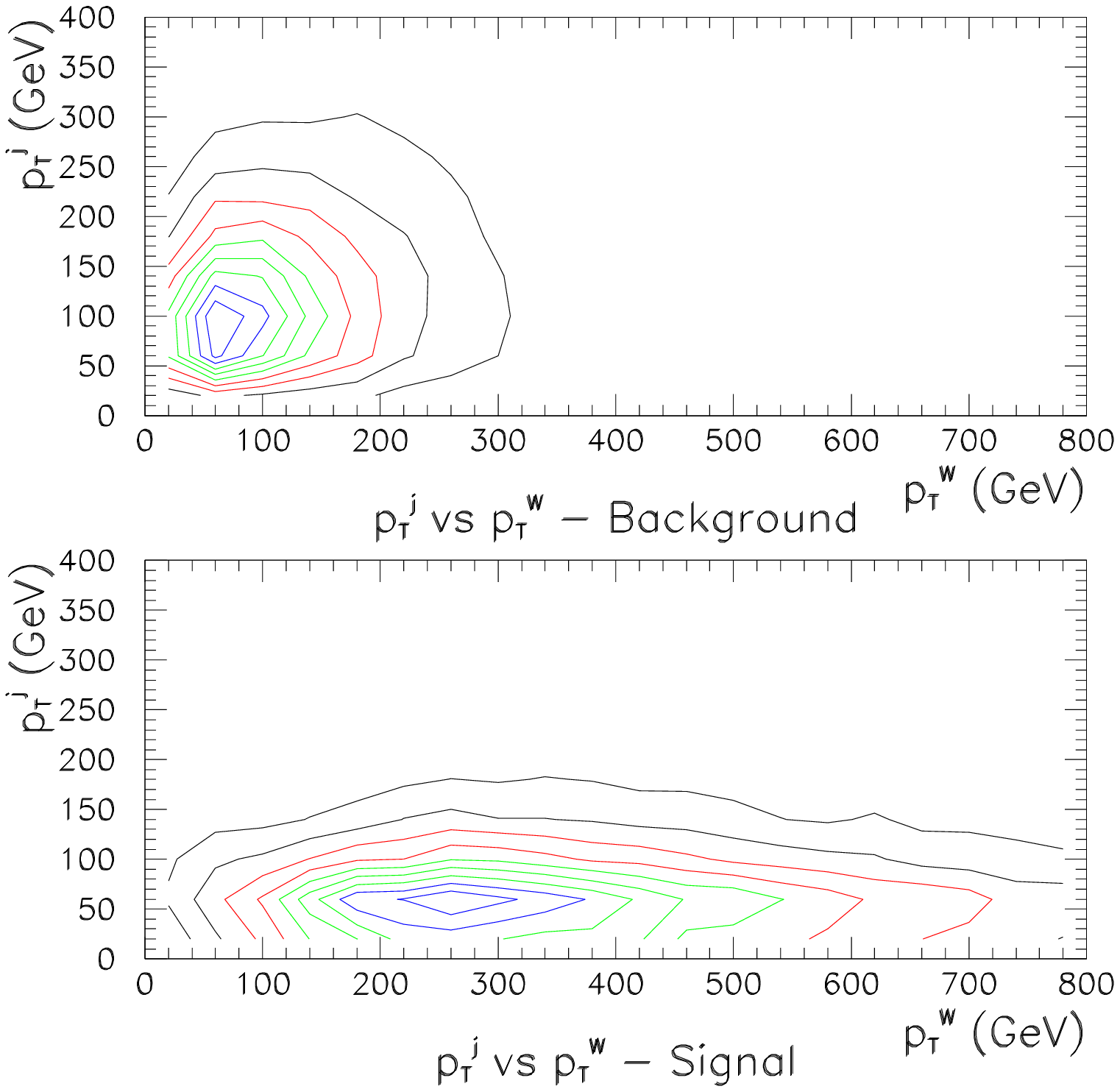,width=0.49\linewidth} &
  \epsfig{file=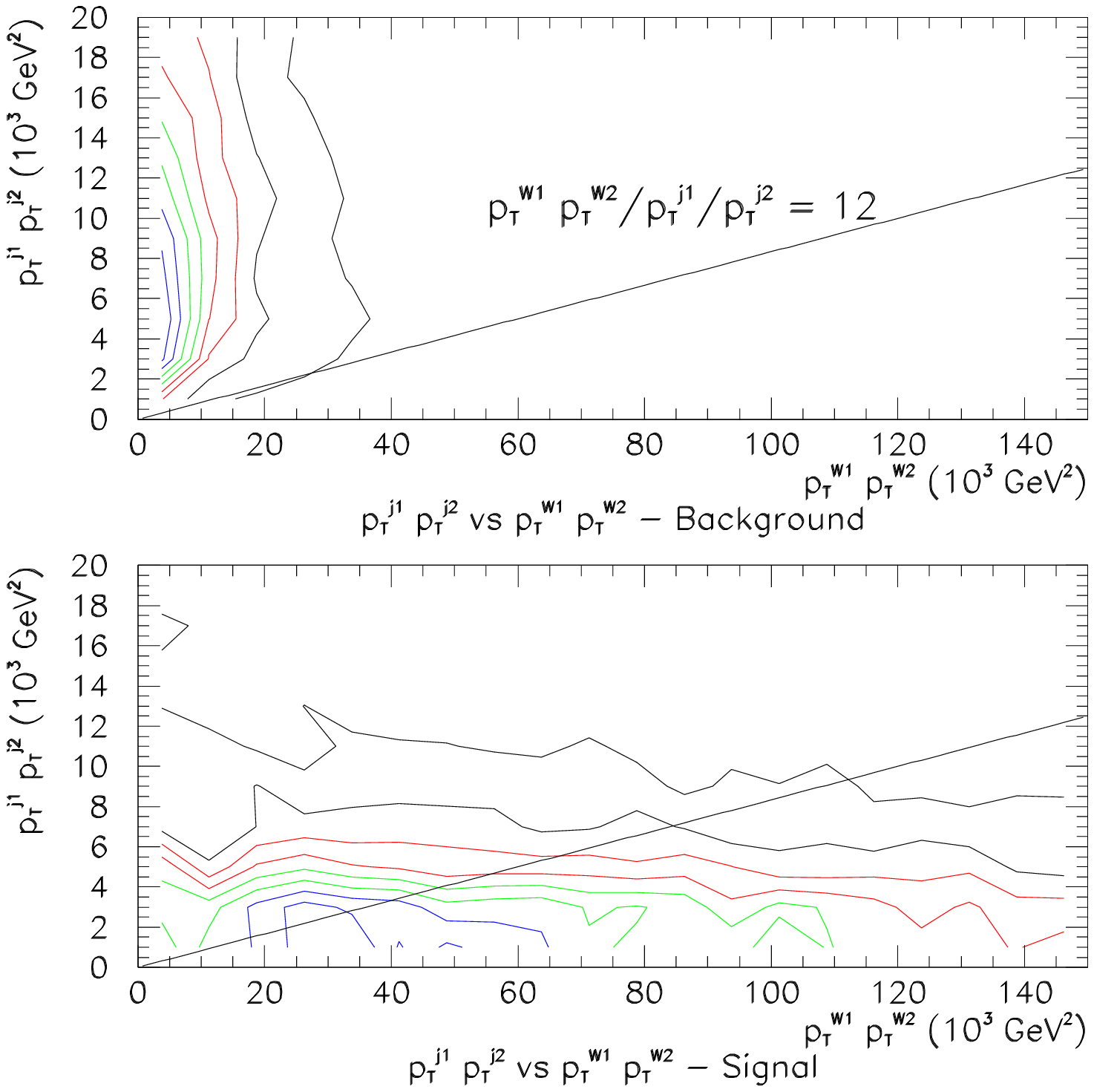,width=0.49\linewidth} \\
  a) & b) \\
  \epsfig{file=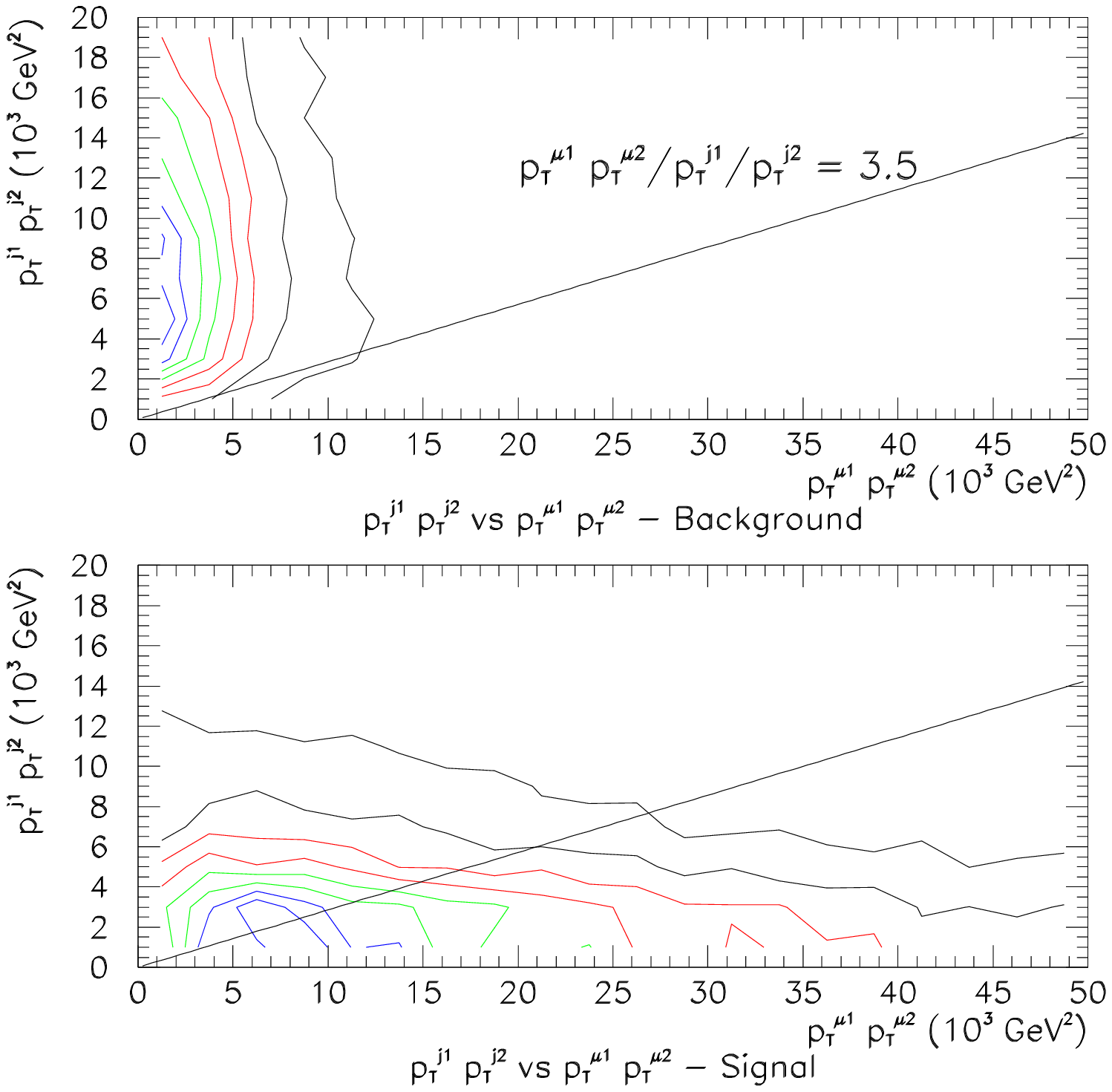,width=0.49\linewidth} &
  \epsfig{file=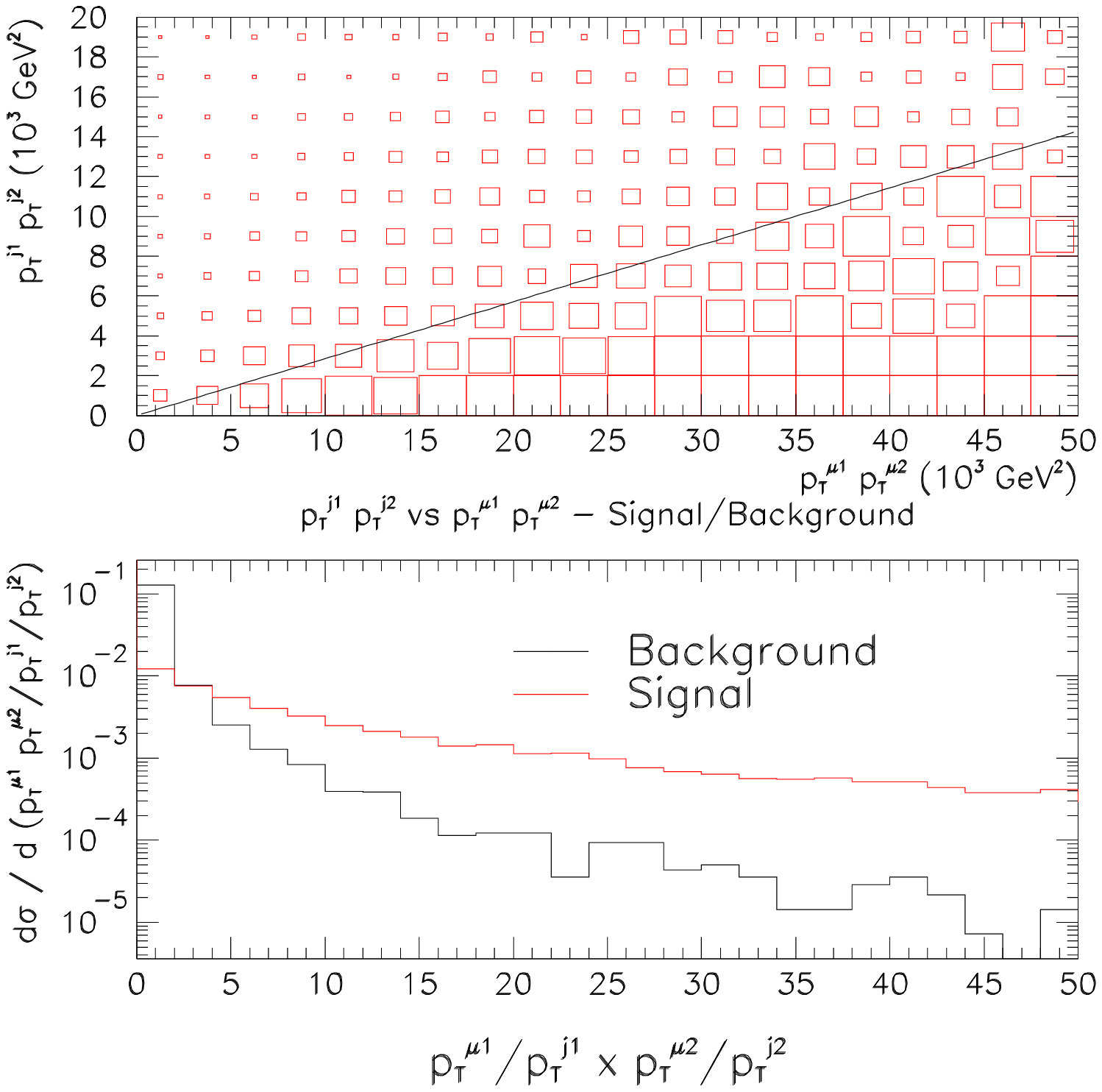,width=0.49\linewidth} \\
  c) & d) 
\end{tabular}
\caption{Transverse momenta and their combinations
  for signal and
  for the irreducible background in the $jjW^+W^+$ channel at 14 TeV,
  after applying the cuts against top production: 
  a) individual $p_T$ of $W$'s and jets,
  b) $p_T$ products of two $W$'s and two jets, 
  c) $p_T$ products of two leptons and two jets, and
  d) the signal to background ratio (top) and the $R_{p_T}$ distribution
  for the signal and for the background (bottom).   Contours in (a) through (c)
represent the lines of constant cross section.}
\label{fig:magic}
\end{figure}

In Section~\ref{sec:prod+em} we proposed a new variable $R_{p_T}$ that
gives a direct signature of a hard $W_LW_L$ scattering process and
that replaces the conventional cuts.  We thus now go back to the point
where only generation level cuts and $t\bar{t}$ cuts have been applied
and study the correlations between the transverse momenta of the $W$
bosons and the jets for those signal and background events which
survived those cuts.  This is shown in Fig.~\ref{fig:magic}(a,b).  It
is apparent that a line corresponding to a constant ratio
\bea
p_T^{~W_1}/p_T^{~j_1} \cdot p_T^{~W_2}/p_T^{~j_2} = 12
\eea
has indeed a large discriminating power between signal and background,
leaving the bulk of the background above, while keeping a substantial
part of the signal below.  In Fig.\,\ref{fig:magic}(c) the same
correlations are shown for the transverse momenta of the outgoing
leptons instead of the $W$'s.  Although $W$ decay tends to smear this
nice picture out (it is of course very important to take proper
account of the correct angular distributions of $W$ decays accordingly
to the different helicities in signal and background, but these
differences unfortunately do not play any positive role in isolating
signal from background), the interesting observation is that the
outgoing leptons from $W$ decay sufficiently keep the basic kinematic
features characteristic for signal and background of their parent
$W$'s.  Consequently, a line of constant ratio
\bea
R_{p_T} = 3.5
\eea
still has a remarkably large discriminating power between signal and
background.  This is illustrated in Fig.~\ref{fig:magic}(d), where the
upper plot shows the ratio of signal to background and in the lower
plot the $R_{p_T}$ distributions for signal and background are given.
When applied on top of all the remaining selection criteria, the
efficiency of a cut that requires $R_{p_T}$ be larger than 3.5 is 0.77
for the signal and 0.14 for the background, which makes it more
effective than all the alternative conventional cuts taken together.

Thus, our complete set of selection criteria is now (set II):
\begin{itemize}
\item 2 same sign leptons,
\item 2 tag jets with $2 < |\eta_j| < 5$ and opposite directions,
\item no $b$-tag,
\item $M_{j_1l_2}$, $M_{j_2l_1} >$ 200 GeV,
\item $M_{jj} >$ 500 GeV,
\item $R_{p_T} >$ 3.5,
\item $\Delta\phi_{ll} >$ 2.5,
\end{itemize}
Note that a combination of only the last two cuts allows to release
several other selection criteria that are conventionally used to cope
with the irreducible background.  The $R_{p_T}$ cut also automatically
removes the $t\bar{t}$ background related to leptonic $B$ decay to a
negligible level, hence we are free to drop any lepton isolation or
central jet veto cut, which may well prove an advantage in a high
pile-up regime like the LHC.  Here we have also modified the $M_{jj}$
cut to a higher value, since we find such change produce an
improvement in combination with the $R_{p_T}$ cut, but not in
combination with the conventional cuts.

\begin{figure}[htbp]
\begin{tabular}{ll}
  \epsfig{file=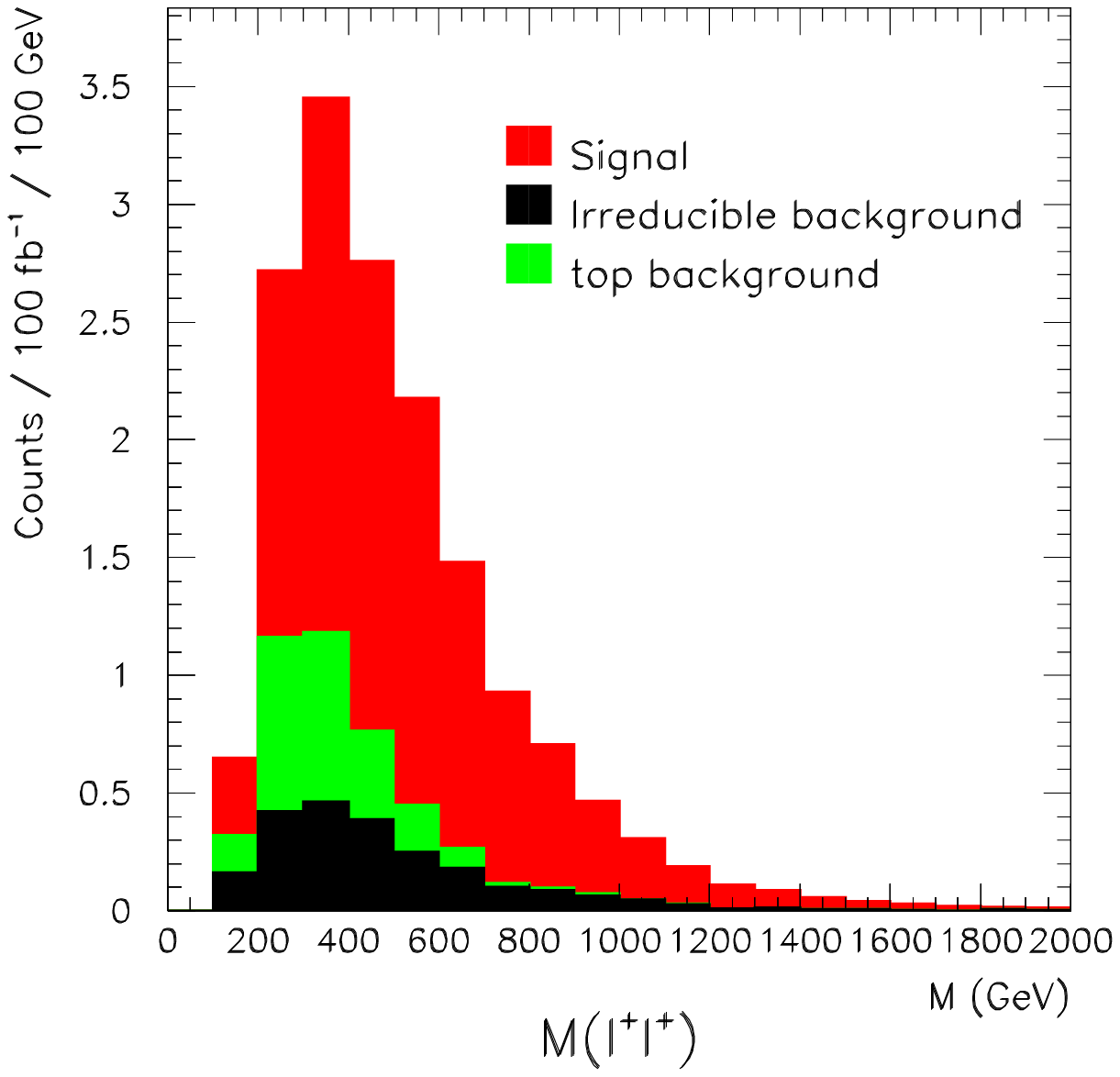,width=0.49\linewidth} &
  \epsfig{file=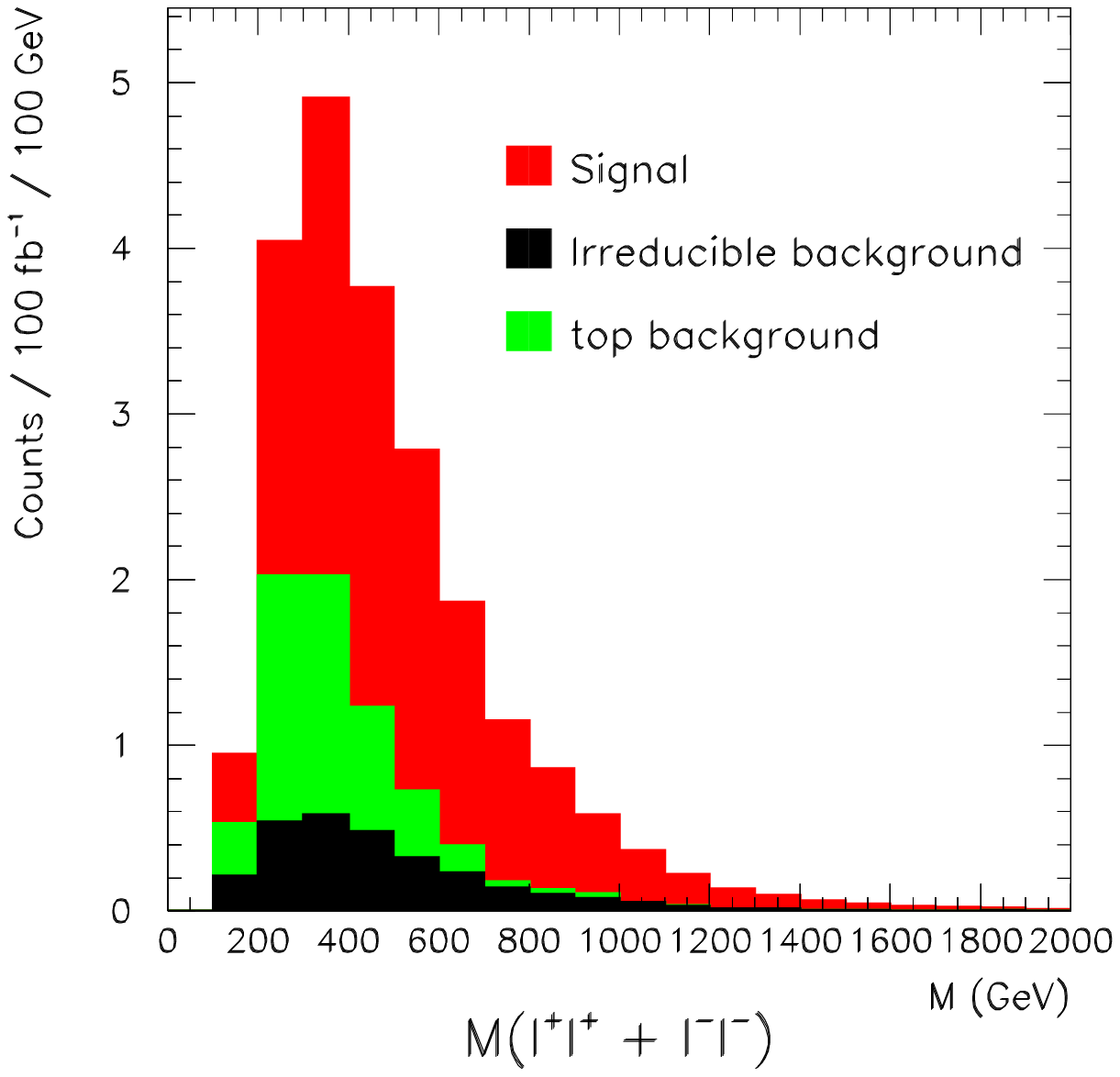,width=0.49\linewidth} \\
  a) & b) 
\end{tabular}
\caption{Invariant mass distributions of two same-sign leptons after
  all the new signal selection criteria (set II), normalized to 100
  $fb^{-1}$: (a) $jjl^+l^+$, (b) $jjl^+l^+$ and $jjl^-l^-$ added.}
\label{fig:results}
\end{figure}

A summary of results obtained with the conventional cuts (set I) and
the $R_{p_T}$ cut (set II) is shown in Tables\,\ref{tab:results} and
\ref{tab:resultsmm}.  For an easy comparison of both results, the
selection criteria were chosen such as to keep in both cases a similar
amount of signal.  Relative to the amount of signal, we find the
$R_{p_T}$ cut improve rejection power against the irreducible
background by a factor 2.5, and against the $t\bar{t}$ background by
about $1/3$.  Fig.\,\ref{fig:results}(a) and (b) show the
corresponding invariant mass distributions of the two same-sign
leptons for this case (to be compared with respective plots in
Fig.\,\ref{fig:results_old}).  We arrive at $S/B \approx 12/4$ once
all the $l^+l^+$ channels are summed up and $S/B \approx 14/6$ by
additionally including the $l^-l^-$ channels.

\begin{table}[htbp]
\begin{center}
{\footnotesize
\begin{tabular}{|c||c|c|c|c|c|c|c|}
  \hline
  Sample & Initial $\sigma$ & Generated & Selected & Selected & Other & Final $\sigma$ & Final $\sigma$ \\
  &  & events & evts (I) & evts (II) & reductions & (I) & (II) \\
  \hline\hline
  $W_L^+W_L^+$ SM       &  7.6 fb &  56485 & 534 & 523 & 0.0426 & 0.0031 fb & 0.0030 fb \\
  $W_L^+W_L^+$ No Higgs & 16.7 fb &  56666 & 11903 & 12313 & {\scriptsize 0.0335 (I)/0.0329 (II)} & 0.1117 fb & 0.1193 fb \\
  Irr. background   & 104.5 fb & 170183 & 1893 & 855 & 0.0426 & 0.0494 fb & 0.0224 fb \\
  $t\bar{t}$ background     &   -    & 15000000 & 1805 & 1318 & 0.00008 & 0.0225 fb & 0.0171 fb \\
  \hline
\end{tabular}
}
\end{center}
\caption{Results of conventional selection criteria (I) and new
  selection criteria (II) for the $jjl^+l^+$ channel.  The signal is
  the result of subtracting the first row from the second row.
  ``Other reductions" stand for losses due to $b$ tagging, lepton
  reconstruction and sign matching, respective branching fractions and
  the unitarity bound, wherever appropriate.}
\label{tab:results}
\end{table}

\begin{table}[htbp]
\begin{center}
{\footnotesize
\begin{tabular}{|c||c|c|c|c|c|c|c|}
\hline
Sample & Initial $\sigma$ & Generated & Selected & Selected & Other & Final $\sigma$ & Final $\sigma$ \\
       &  & events & evts (I) & evts (II) & reductions & (I) & (II) \\
\hline\hline
$W_L^-W_L^-$ SM       & 1.72 fb &   9078 &  93 &  94 & 0.0426 & 0.0008 fb & 0.0008 fb \\
$W_L^-W_L^-$ No Higgs & 4.30 fb &   9298 &  1579 &  1747 & {\scriptsize 0.0360 (I)/0.0357 (II)} & 0.0263 fb & 0.0288 fb \\
Irr. background   & 20.35 fb &  35615 &  525 & 241 & 0.0426 & 0.0128 fb & 0.0059 fb \\
$t\bar{t}$  background  &   -    & 15000000 & 1805 & 1318 & 0.00008 & 0.0225 fb & 0.0171 fb \\
\hline
\end{tabular}
}
\end{center}
\caption{Results of conventional selection criteria (I) and new
  selection criteria (II) for the $jjl^-l^-$ channel.  The signal is
  the result of subtracting the first row from the second row.
  ``Other reductions" stand for losses due to $b$ tagging, lepton
  reconstruction and sign matching, respective branching fractions and
  the unitarity bound, wherever appropriate.}
\label{tab:resultsmm}
\end{table}

It is interesting to note that $R_{p_T}$ being a dimensionless number
does not depend on $\sqrt{s}$ to first approximation.  In principle,
exactly the same cut could be also used in an analysis of $pp$ data at
7 TeV or at 8 TeV.  The efficiency of such cut (calculated on top of
all the other cuts) at $\sqrt{s} = 7$ TeV (8 TeV) is 0.71 (0.72) for
the signal and 0.12 (0.12) for the irreducible background, which is
only slightly weaker than for $\sqrt{s} = 14$ TeV.  It is mainly due
to the much lower signal cross section to begin with (factor $\sim$7
at the level of generation cuts) that 7 or 8 TeV ultimately does not
offer the possibility to observe the signal in a viable amount of
time.  However, the efficiency of the $R_{p_T}$ variable can be
experimentally tested at 7 or 8 TeV.

\section{The opposite-sign $WW$ channel}
\label{sec:oppww}

\begin{figure}[htbp]
\begin{tabular}{ll}
  \epsfig{file=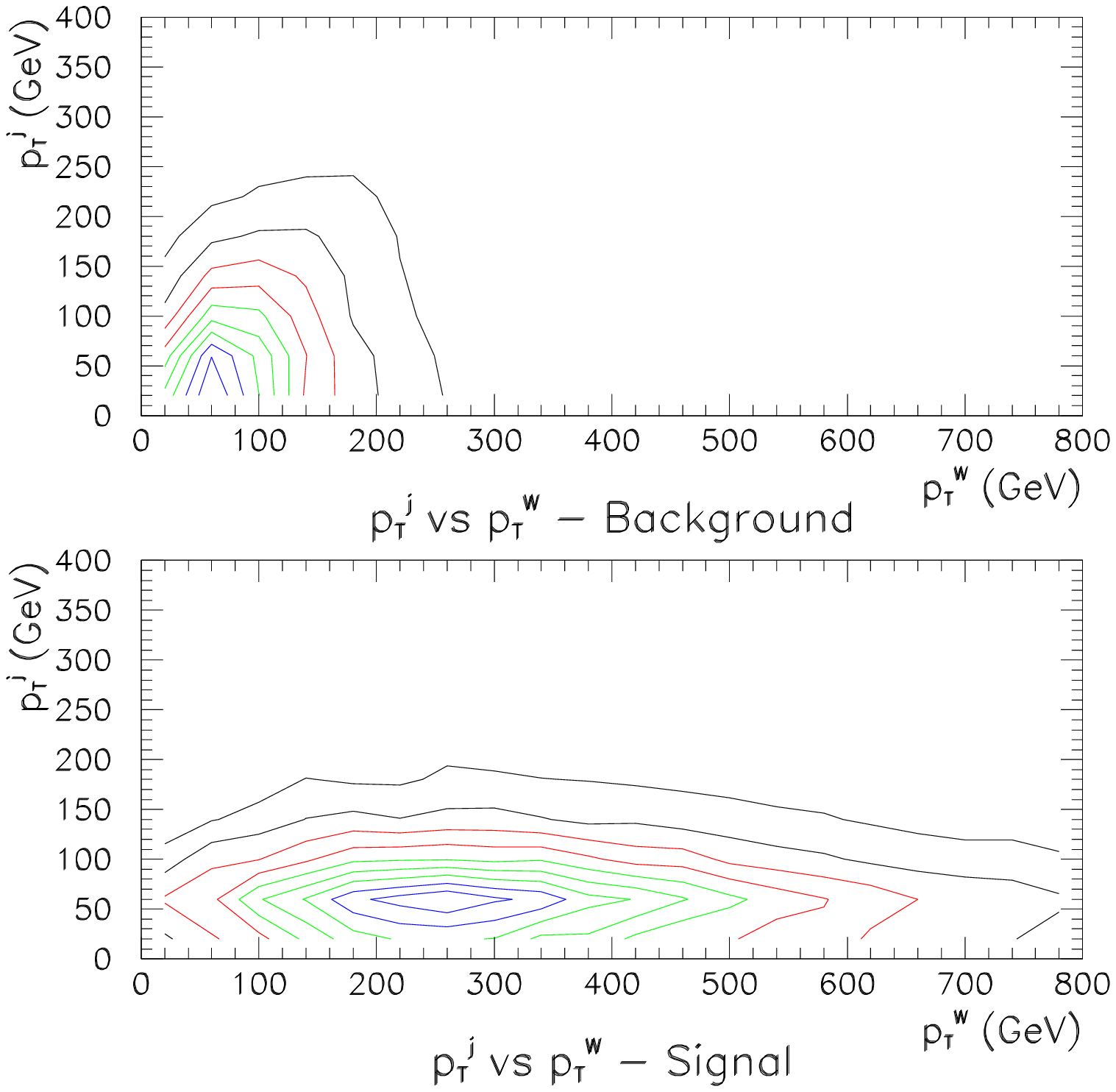,width=0.49\linewidth} &
  \epsfig{file=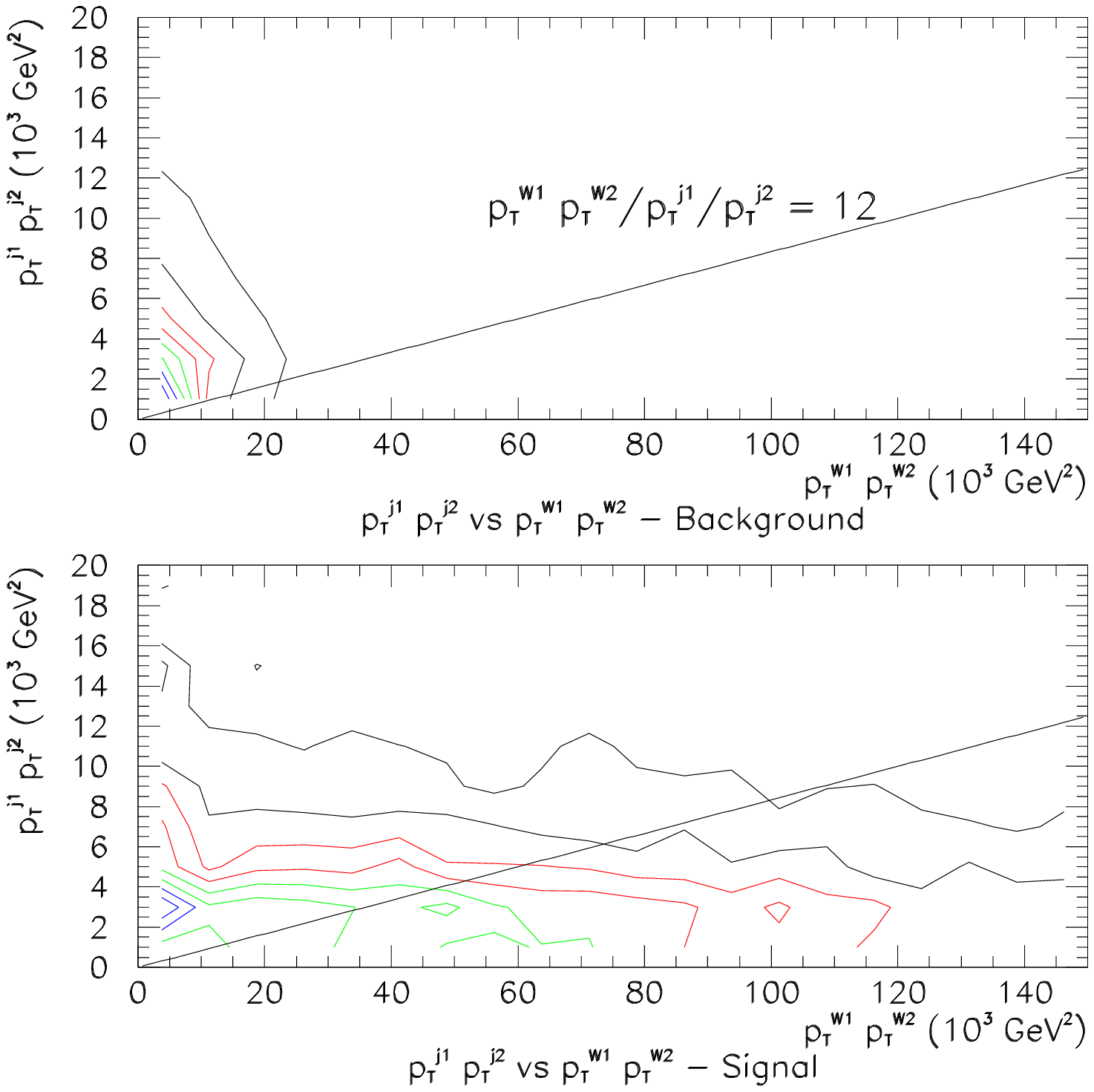,width=0.49\linewidth} \\
  a) & b) \\
  \epsfig{file=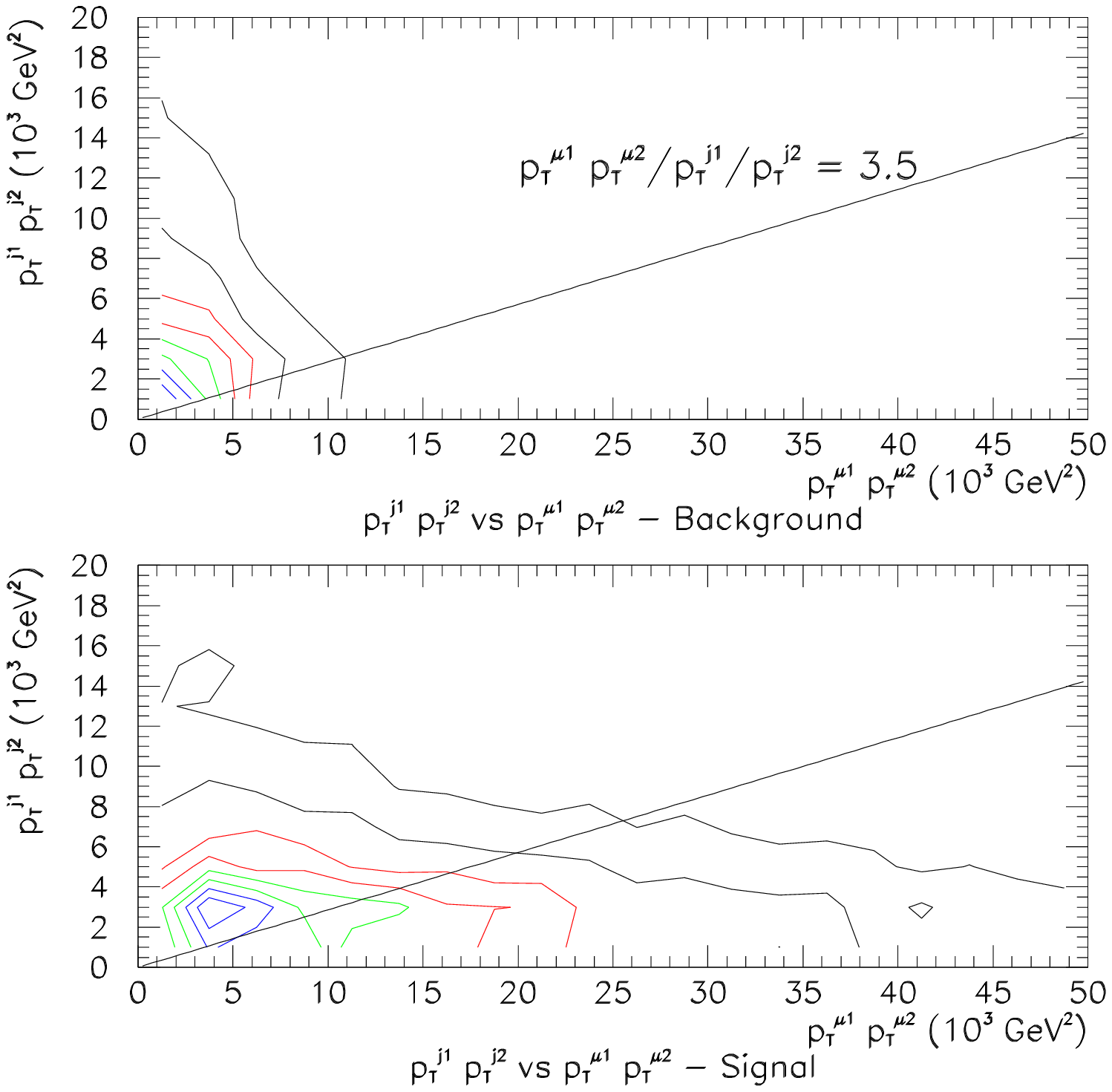,width=0.49\linewidth} &
  \epsfig{file=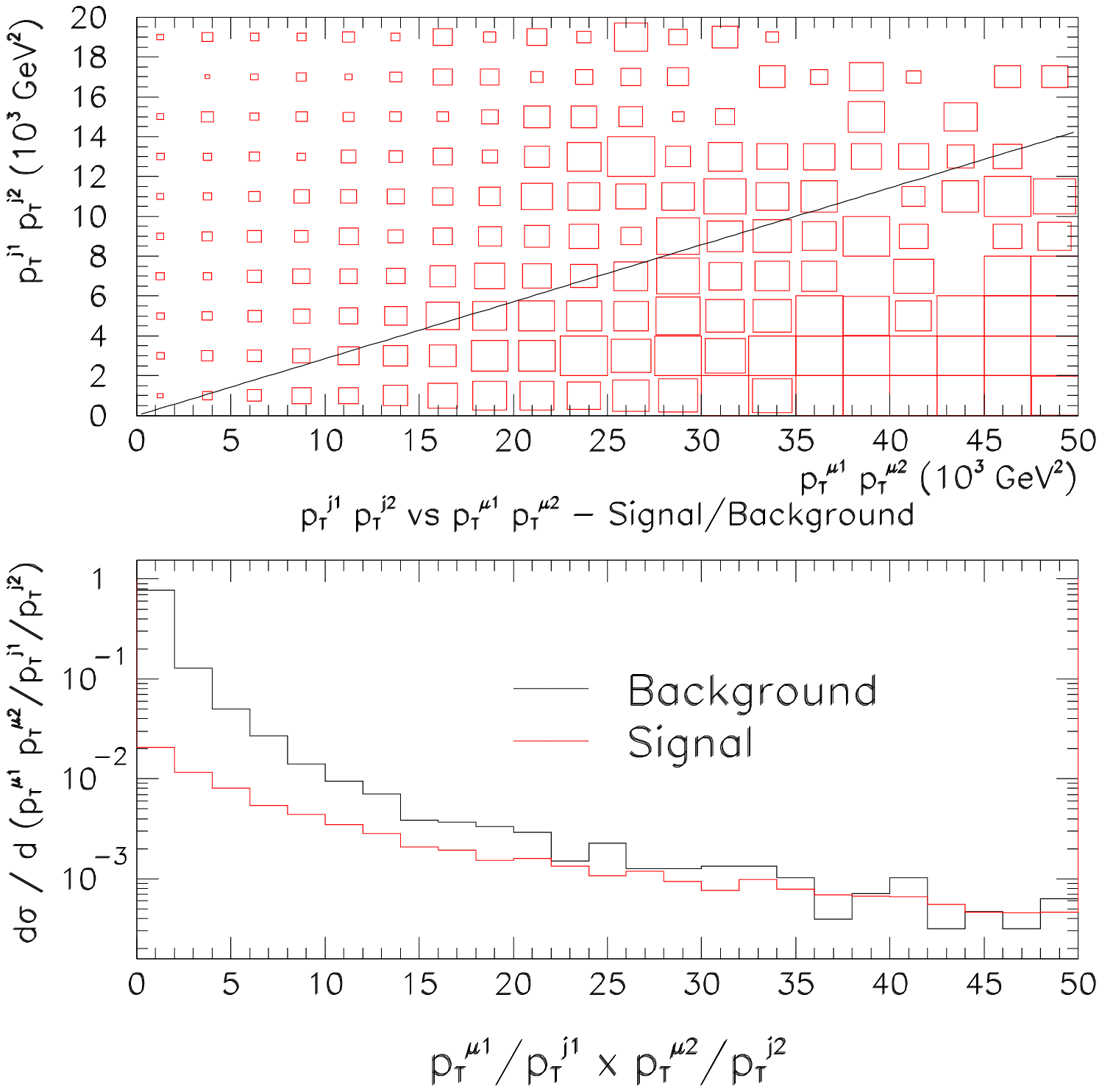,width=0.49\linewidth} \\
  c)  & d) 
\end{tabular}
\caption{Transverse momenta and their combinations for signal and
  for the irreducible background in the $jjW^+W^-$ channel at 14 TeV,
  after applying the cuts against top production:
  a) individual $p_T$ of $W$'s and jets,
  b) $p_T$ products of two $W$'s and two jets,   
  c) $p_T$ products of two leptons and two jets, and
  d) the signal to background ratio (top) and the $R_{p_T}$ distribution
  for the signal and for the background (bottom).  Contours in (a) through (c)
represent the lines of constant cross section.}
\label{fig:magicpm}
\end{figure}

As already stated in Section~\ref{sec:prod+em}, due to the many
additional contributions to the background in the $jjW^+W^-$ channel
(both in terms of Feynman diagrams and physical processes), there is
good reason to expect a lower $R_{p_T}$ efficiency.  In this Section
we present our own analysis of the $jjW^+W^-$ channel for the sake of
a direct comparison.  Generated samples, initial and final cross
sections for the signal and for the backgrounds are shown in
Table~\ref{tab:resultspm}.  Cuts imposed at the generation level were:
\begin{itemize}
\item $|\eta_j| <$ 5,
\item $\Delta\eta_{jj} >$ 4,
\item $p_T^{~j} >$ 10 GeV,
\item $M_{jj} >$ 350 GeV.
\end{itemize}
An attempt to exploit the kinematic correlations between $W$'s and
jets and between leptons and jets, in a similar manner as was applied
for the same-sign channel and presented in Section~\ref{sec:sameww},
is shown in Fig.~\ref{fig:magicpm}.  Many important differences with
respect to the same-sign channel are readily visible.  First and
foremost, background extends to much lower jet transverse momenta, to
the effect of background and signal being marginally distinguishable
in the jet $p_T$ distributions (Figs.~\ref{fig:magicpm}(a) and (b)).
In these circumstances, the $R_{p_T}$ ratio can only be as effective
in separating signal from background as lepton transverse momenta on
their own.  Fig.~\ref{fig:magicpm}(d) confirms that no particular cut
on $R_{p_T}$ can bring background levels below or even close to the
signal level without risking very low efficiency.  This study
concludes that $R_{p_T}$ cannot be used as an effective discriminant.
Larger initial cross sections for the signal are unfortunately coupled
with a weaker kinematic separation from the background in $p_T^{~l}$
(see Fig.~\ref{fig:magicpm}(c)).  However, we find the sum $p_T^{~l_1}
+ p_T^{~l_2}$ or the product $p_T^{~l_1}\cdot p_T^{~l_2}$ still have a
slightly larger efficiency than combined cuts on the individual
$p_T^{~l}$.  Another important disadvantage is that top production can
directly fake the signal signature here, rather than being restricted
to small experimental effects, and an additional central jet veto is
necessary to bring this background down to a manageable level.  A veto
on additional jets with $p_T > 25$ GeV and $\eta$ anywhere between
$\eta_{j1}$ and $\eta_{j2}$ removes nearly 75\% of the surviving
$t\bar{t}$ events at the expense of an additional 15\% of the signal.

As usual, we require no tagged $b$ and no additional leptons with $p_T
> 10$ GeV within $|\eta| <$ 2.  Finally, our full selection criteria
are:
\begin{itemize}
\item exactly 2 opposite sign leptons within detector acceptance,
\item 2 tag jets with $2 < |\eta_j| < 5$ and opposite directions,
\item no $b$-tag,
\item $M_{j_1l_2}$, $M_{j_2l_1} >$ 200 GeV,
\item $M_{jj} >$ 500 GeV,
\item central jet veto,
\item $p_T^{~l_1} + p_T^{~l_2} >$ 300 GeV,
\item $\Delta\phi_{ll} >$ 2.5,
\item $M_{ll} >$ 300 GeV.
\end{itemize}

\begin{table}[htbp]
\begin{center}
{\small
\begin{tabular}{|c||c|c|c|c|c|c|c|}
\hline
Sample &   Initial $\sigma$ & Generated & Selected & Other      & Final $\sigma$ \\
       &                    & events    & events   & reductions &   \\
\hline\hline
$W_L^+W_L^-$ EW SM &       24.9 fb &  90296 &      777 &     0.0426 & 0.0091 fb \\
$W_L^+W_L^-$ EW No Higgs & 44.0 fb &  102570 &     8257 &     0.0362 & 0.1282 fb \\
Irr. background  &  3.37 pb &    500000 &      458 &     0.0426 & 0.1315 fb \\
$t\bar{t}$ background & -     &  15000000 &      194 &     0.0028 & 0.3610 fb \\
\hline
\end{tabular}
}
\end{center}
\caption{Results for the $jjl^+l^-$ channel.  The signal is the result
  of subtracting the first row from the second row.  ``Other
  reductions" stand for losses due to $b$ tagging, lepton
  reconstruction, respective branching fractions and the unitarity
  bound, wherever appropriate.}
\label{tab:resultspm}
\end{table}

The result (see Table~\ref{tab:resultspm}) is consistent with those of
previously published analyses on the same channel~\cite{ballestrero,
  zeppenfeld} (taking into account all the differences in the
respective analysis approaches), but does not bring much improvement
to the subject.

\section{Discussion and outlook}
\label{sec:summary}

We have presented a detailed phenomenological analysis of $WW$
scattering at the LHC, which in some aspects contains improvements
with respect to the analyses published by other authors.  Our analysis
was optimized primarily for the discriminating power between
alternative scenarios of EWSB and unitarity violation.  The
discrimination is based essentially on a counting experiment in which
a light SM-like Higgs boson as the only source of EWSB should ideally
produce an almost null result.

It is not the aim of this work to fine tune the selection in order to
produce the best possible $S/B$, as the exact criteria will change
anyway once a full detector simulation is included.  Regardless of
that it seems clear that the $R_{p_T}$ variable is able to provide
some new insight into the subject and possibly improve the expected
performance in the same-sign $WW$ channel.  Our result with
conventional cuts is roughly consistent with previously published
results of similar analyses, in particular with the recent
Ref.~\cite{ballestrero}, if the same conditions and assumptions are
applied in both cases.  Their best result $S/B = 13/6$ is the
equivalent of our $S/B \approx 17/7$ per 100 $fb^{-1}$.  In fact, most
of the difference comes from the fact that their analysis includes a
cut on the minimum jet $p_T$ at 30 GeV which is the standard in LHC
experiments, but very costly to the signal as far as $W_LW_L$ selection
at high invariant mass is concerned.  With the $R_{p_T}$ cut
instead this result could become even $S/B \approx 17/3$.

The question of a practical minimum jet $p_T$ cut, in particular
having in view the large LHC pile-up, is
an issue that requires a careful dedicated study which is clearly
beyond the scope of the present work.  In this paper we show a strong
physics motivation for releasing the stringent cuts on jet $p_T$
(our analysis implicitly assumes $p_T > 10$ GeV, i.e., jets below
that threshold are not reconstructed).  The idea of tagging only one 
forward jet with $p_T > 30$ GeV, which should be rather safe, and working 
out additional criteria to indentify correctly its softer companion,
is an interesting possibility that needs to be investigated.
Data collected in 2012 may already shed some light on the issue.

Our new selection criterion provides potential to improve the final
figure of merit regardless of any working assumptions or
approximations we have done throughout this paper.  It is clear,
however, that many important unknowns still remain and will require a
careful experimental study.  One is the vulnerability of $R_{p_T}$ to
experimental smearing in the measurement of all the individual
$p_T$'s.  Since $R_{p_T}$ is expected to work for 7 or 8 TeV nearly as
efficiently as for 14 TeV, the practical effects of $p_T$ smearing can
be studied at the LHC on 2012 data and reasonably assumed not to be
worse for future 14 TeV data.

Top background poses more unknowns.  The total cross section for top
production at 14 TeV will ultimately have to be measured.  Since
$t\bar{t}$ background to the $W^+W^+$ channel arising from lepton sign
misidentification is not completely negligible, the practical profits
from using the $R_{p_T}$ variable will depend on precise knowledge
of the sign matching efficiencies of the individual detectors
(for demonstration purposes we could of course tune the
selection so to kill $t\bar{t}$ completely, but such procedure would
also end up in very low signal and hence is of little practical
interest).  One should note that an important cross check of the
$t\bar{t}$ background evaluation can be obtained by comparing the
results of the two in principle separate measurements that this
analysis involves: the $jjl^+l^+$ channel and the $jjl^-l^-$ channel.
Since the relative amounts of $WW$ in the two channels are like 4:1
and the amounts of $t\bar{t}$ like 1:1, the actual ratio will shed
some light on the true composition of the sample.  With enough data
having been collected, the $jjl^-l^-$ channel could be even used to
estimate the remaining $t\bar{t}$ background and subtract it.

Certainly it will be crucial to include the electron channels in
addition to the experimentally purest muon channels, as the latter
alone do not guarantee enough statistical significance.  A further,
more detailed, study based on a full detector simulation and
detector-specific event reconstruction is vital for the good
understanding of rejection capabilities of the detector backgrounds
and the resulting purity of electron reconstruction.

Overall, results presented in this paper show that there is good
chance to get some important hints about the mechanism of electroweak
symmetry breaking already after the first $100 fb^{-1}$ of data at 14
TeV, by focusing on the purely leptonic $W$ decay channels.  We
advocate the importance of the same-sign $WW$ channel as the most
promising one, particularly in the absence of low mass resonances (1-2
TeV) in the $W^+W^-$ channel.

\section*{Acknowledgments}

This work has been supported in part by the Polish Ministry of Science and
Higher Education as research projects 666/N-CERN/2010/0, N N202 230337 (2009-12)
and N N202 103838 (2010-12) and by the Collaborative Research Center
SFB676/1-2006 of the DFG.

\end{document}